\documentclass[reqno,natbib,runningheads,11pt]{amsart}%
\usepackage{graphicx}
\usepackage{amssymb}
\usepackage{amsfonts}
\usepackage{amsmath}
\usepackage{lscape}
\usepackage{graphicx}
\usepackage{hyperref}
\usepackage{lineno}%
\providecommand{\U}[1]{\protect\rule{.1in}{.1in}}
\newtheorem{theorem}{Theorem}[section]

\newtheorem{remark}{Remark}[section]

\begin{document}

\begin{center}
\medskip{\Large {\textbf{Copula representation of bivariate \textit{L}%
-moments: }}}

\medskip{\Large {\textbf{A new estimation method for multiparameter }}}

{\Large {\textbf{2-dimentional copula models}}} \bigskip

{\large \textbf{Brahim Brahimi}$^{\text{a}}$, \textbf{Fateh Chebana}%
$^{\text{b}}$, \textbf{and Abdelhakim Necir}$^{\text{a},*}$}\bigskip

{\small {$^{\text{a}}$\textit{Laboratory of Applied Mathematics, Mohamed
Khider University of Biskra, \break Biskra 07000, Algeria}}\smallskip}

{\small { $^{\text{b}}$\textit{Canada Research Chair on the Estimation of
Hydrometeorological Variables, \break INRS-ETE, 490, rue de la Couronne, G1K
9A9 Quebec (QC), Canada}} }
\end{center}

\vspace{2cm} \bigskip\noindent\rule{\linewidth}{0.2mm} \noindent
\textbf{Abstract.} \medskip

\noindent Recently, Serfling and Xiao (2007) extended the \textit{L}-moment
theory\ (Hosking, 1990) to the multivariate setting. In the present paper, we
focus on the two-dimension random vectors to establish a link between the
bivariate \textit{L}-moments (BLM) and the underlying bivariate copula
functions. This connection provides a new estimate of dependence parameters of
bivariate statistical data. Consistency and asymptotic normality of the
proposed estimator are established. Extensive simulation study is carried out
to compare estimators based on the BLM, the maximum likelihood, the minimum
distance and rank approximate \textit{Z}-estimation. The obtained results show
that, when the sample size increases, BLM-based estimation performs better as
far as the bias and computation time are concerned. Moreover, the root mean
squared error\ (RMSE) is quite reasonable and less sensitive in general to
outliers than those of the above cited methods. Further, we expect that the
BLM method is an easy-to-use tool for the estimation of multiparameter copula
models.\bigskip

 \medskip


\noindent\rule{\linewidth}{0.2mm} \vfill

\noindent{\small $^{\text{*}}$Corresponding author: (A.~Necir):
\texttt{necirabdelhakim@yahoo.fr} \newline\texttt{brah.brahim@gmail.com}
(B.~Brahimi), \newline\texttt{Fateh.Chebana@ete.inrs.ca} (F.~Chebana)}


\section{\textbf{Introduction and motivation}}

\noindent The copula method is a tool to construct multivariate distributions
and describe the dependence structure in multivariate data sets (e.g., Joe,
1997 or Nelsen, 2006). Modelling dependence structures by copulas is a topic
of current research and of recent use in several areas, such as financial
assessments (e.g., Malevergne and Sornette, 2003), insurance (e.g., Drees and
M\"{u}ller, 2008) and hydrology (e.g., Dupuis, 2007). For the sake of
simplicity, throughout the paper, we restrict ourself to the two-dimensional
case. Let $\left(  X^{\left(  1\right)  },X^{\left(  2\right)  }\right)  $ be
a bivariate random variable with joint distribution function (df)
\[
F\left(  x_{1},x_{2}\right)  =\mathbb{P}\left(  X^{\left(  1\right)  }\leq
x_{1},X^{\left(  2\right)  }\leq x_{2}\right)  ,\text{ }\left(  x_{1}%
,x_{2}\right)  \in\mathbb{R}^{2},
\]
and marginal df $F_{j}\left(  x_{j}\right)  =\mathbb{P}\left(  X^{\left(
j\right)  }\leq x_{j}\right)  $ for $x_{j}\in\mathbb{R}$ and $j=1,2.$ If not
stated otherwise, we assume that the $F_{j}$ are continuous functions.
According to Sklar's theorem (Sklar, 1959) there exists a unique copula
$C:\mathbb{I}^{2}\rightarrow\mathbb{I},$ with $\mathbb{I=}\left[  0,1\right]
,$ such that%
\[
F\left(  x_{1},x_{2}\right)  =C\left(  F_{1}\left(  x_{1}\right)
,F_{2}\left(  x_{2}\right)  \right)  ,\text{ for }\left(  x_{1},x_{2}\right)
\in\mathbb{R}^{2}.
\]
The copula $C$ is the joint df of the uniform random variables (r.v.'s)
$U_{j}=F_{j}\left(  X^{\left(  j\right)  }\right)  ,$ $j=1,2,$ defined for
$\left(  u_{1},u_{2}\right)  \in\mathbb{I}^{2},$ by
\[
C\left(  u_{1},u_{2}\right)  =F\left(  F_{1}^{-1}\left(  u_{1}\right)
,F_{2}^{-1}\left(  u_{2}\right)  \right)  ,
\]
where $G^{-1}$ is the generalized inverse function (or the quantile function)
of a df $G.$\medskip

\noindent A parametric copula model arises for $\left(  X^{\left(  1\right)
},X^{\left(  2\right)  }\right)  $ when $C$ is unknown but assumed to belong
to a class $\mathcal{C}:=\left\{  C_{\mathbf{\theta}},\text{ }\mathbf{\theta
}\in\mathcal{O}\right\}  ,$ where $\mathcal{O}$ is an open subset of
$\mathbb{R}^{r}$ for some integer $r\geq1.$ Statistical inference on the
dependence parameter $\mathbf{\theta}$ is one of the main topics in
multivariate statistical analysis. Several methods of copula parameter
estimation have been developed, including the pseudo maximum likelihood (PML),
inference of margins, minimum distance and others, see for instance Genest
\textit{et al.}\ (2009). All these methods use constrained some optimization
techniques, and may require substantial computational time. In this paper, we
present a new estimation method of $\theta$ based on the bivariate
\textit{L}-moments that serves as alternative in front of computation's time
issue and produces results leads to reasonable estimation performances. The
multivariate \textit{L}-moments have been introduced by Serfling and Xiao
(2007) as an extension of the univariate \textit{L}-moments introduced by
Hosking (1990). The \textit{L}-comoments have interpretations similar to the
classical central moment covariance, coskewness, and cokurtosis that also
possess the features of the \textit{L}-moments. This extension is useful to
solve some problems in connection with multivariate heavy-tailed distributions
and small samples. As mentioned, for instance, in Hosking (1990) and recently
in Delicado and Goria (2008), the main advantage of \textit{L}-moments
vis-a-vis of classical estimation methods (e.g. least squares, moments and
maximum likelihood) is their relative slight sensitivity to outlying data and
their performance in statistical inference with small samples.\medskip

\noindent In this paper we establish a functional representation of bivariate
\textit{L}-moments (BLM) by the underlying copula function and propose a new
estimation method for parametric copula models. By considering multiparameter
Farlie-Gumbel-Morgenstern (FGM) and Archimedean copulas, simulation studies
are carried out to compare the performance of this method with those of the
PML, minimum distance (MD) and rank approximate \textit{Z}-estimation.\medskip

\noindent The L-moment approach is of interest for multiparameter
distributions. In the case of one parameter distributions, it is equivalent to
the classical method of moments (see e.g. Hosking, 1990)(since the first
L-moment corresponds to the expected value). Next we see, that this property
applies also in the case of multivariate setting for multiparameter
copulas.\medskip

\noindent The rest of the paper is organized as follows. In Section
\ref{section2}, we briefly introduce the univariate and bivariate
\textit{L}-moment approaches. We present, in Section \ref{section3},
functional representations of the bivariate \textit{L}-moments by copula
functions and give some examples. A new estimator of copula parameter and its
asymptotic behavior are given in Sections \ref{section4}. In Section
\ref{simulation}, a simulation study evaluates the BLM performance is given.

\section{\label{section2}\textbf{Bivariate \textit{L}-moments}}

\noindent First we begin with a brief introduction on the univariate
\textit{L}-moments. Hosking (1990) introduced \textit{L}-moments $\lambda_{k}$
as an alternative to the classical central moments $\mu_{k}=\mathbb{E}\left[
\left(  Y-\mu\right)  ^{k}\right]  $ determined by the df $F_{Y}$ of the
underlying r.v. $Y.$ An \textit{L}-moment $\lambda_{k}$ is defined as a
specific linear combination of the expectations of the order statistics
$Y_{1:k}\leq...\leq Y_{k:k}.$ More precisely, the $k$th \textit{L}-moment is
defined by
\[
\lambda_{k}=\frac{1}{k}\sum_{\ell=0}^{k-1}\frac{\left(  -1\right)  ^{\ell
}\left(  k-1\right)  !}{\ell!\left(  k-1-\ell\right)  !}\mathbb{E}\left[
Y_{k-\ell:k}\right]  ,\text{ }k=1,2,...
\]
By analogy with the classical moments, the first four \textit{L}-moments
$\lambda_{1},$ $\lambda_{2},$ $\lambda_{3}$ and $\lambda_{4}$ measure
location, scale, skewness and kurtosis features respectively. The
\textit{L}-functional representation of $\lambda_{k}$ is terms of the quantile
function $F_{Y}^{-1}$ is given by (see Hosking, 1998):
\begin{equation}
\lambda_{k}=\int_{\mathbb{I}}F_{Y}^{-1}\left(  u\right)  P_{k-1}\left(
u\right)  du, \label{eq1}%
\end{equation}
where $P_{k}\left(  u\right)  :=\sum\limits_{\ell=0}^{k}p_{k,\ell}u^{\ell},$
with $p_{k,\ell}=\left(  -1\right)  ^{k+\ell}\left(  k+\ell\right)  !/\left[
\left(  \ell^{2}\right)  !\left(  k-\ell\right)  !\right]  $ is the shifted
Legendre polynomials (SLP). In the sequel, we will make use of the three first
SLP%
\[
P_{0}\left(  u\right)  =1,\text{ }P_{1}\left(  u\right)  =2u-1,\text{ }%
P_{2}\left(  u\right)  =6u^{2}-6u+1.
\]
A straightforward transformation in $\left(  \ref{eq1}\right)  $ using
$P_{0}\equiv1$ and the orthogonality of $P_{k-1}$ leads to a representation in
terms of covariance, that is%
\begin{equation}
\lambda_{k}=\left\{
\begin{array}
[c]{ll}%
\mathbb{E}\left[  Y\right]  & k=1;\medskip\\
Cov\left(  Y,P_{k-1}\left(  F_{Y}\left(  Y\right)  \right)  \right)  &
k=2,3,...
\end{array}
\right.  \label{eq2}%
\end{equation}

\noindent\textit{L}-moments may be used as summary statistics for data
samples, to identify probability distributions and fit them to data. A brief
description of these methods is given in Hosking (1998). \textit{L}-moments
are now widely used in water sciences especially in flood frequency analysis.
Recent studies include Kjeldsen \textit{et al.} (2002), Kroll and Vogel
(2002), Lim and Lye (2003), Chebana and Ouarda (2007) and Chebana \textit{et
al.} (2009). In other recent work, Karvanen \textit{et al.} (2002) used
\textit{L}-moments for fitting distributions in independent component analysis
in signal processing, and Jones and Balakrishnan (2002) pointed out some
relationships between integrals occurring in the definition of moments and
\textit{L}-moments. Hosking (2006) showed that, for a wide range of
distributions, the characterization of a distribution by its \textit{L}%
-moments is non-redundant. That is, if one \textit{L}-moment is dropped, the
remaining \textit{L}-moments no longer suffice to determine the entire
distribution. Recently, Serfling and Xiao (2007) extended this approach to the
multivariate case, this has already begun to be developed and applied in
statistical hydrology by Chebana and Ouarda (2007) and Chebana \textit{et al.}
(2009).$\medskip$

\noindent Next we present basic notations and definitions of the bivariate
\textit{L}-moments. Let $X^{\left(  1\right)  }$ and $X^{\left(  2\right)  }$
be two r.v.'s with finite means, margins $F_{1}$ and $F_{2}$ and
\textit{L}-moments sequences $\lambda_{k}^{\left(  1\right)  }$ and
$\lambda_{k}^{\left(  2\right)  },$ respectively. By analogy with the
covariance representation $\left(  \ref{eq2}\right)  $ for \textit{L}-moments,
and the central comoments, Serfling and Xiao (2007) defined the $k$th
\textit{L}-comoment of $X^{\left(  1\right)  }$ with respect to $X^{\left(
2\right)  }$ by the covariance of the couple of r.v.'s $X^{(1)}$ and
$P_{k-1}(F_{2}(X^{(2)})),$ for every $k\geq2,$ as%
\[
\lambda_{k[12]}=Cov\left(  X^{(1)},P_{k-1}\left(  F_{2}\left(  X^{(2)}\right)
\right)  \right)  .
\]
\noindent Thus, the $k$th \textit{L}-comoment of $X^{\left(  2\right)  }$ with
respect to $X^{\left(  1\right)  }$ is defined by
\[
\lambda_{k[21]}=Cov\left(  X^{(2)},P_{k-1}\left(  F_{1}\left(  X^{(1)}\right)
\right)  \right)  .
\]
\noindent If we suppose that $F$ belongs to a parametric family of df's, then
the $k$th \textit{L}-comoment $\lambda_{k[12]}$ depends on the parameters
relies the margins and the dependence structure between $X^{(1)}$ and
$X^{(2)}.$ Since we focus only on the estimation of copula parameters, then it
is convenient to use the $k$th \textit{L}-comoment of $F_{1}\left(  X^{\left(
1\right)  }\right)  $ with respect to $X^{\left(  2\right)  }$ instead of
$\lambda_{k[12]},$ that is%
\[
\lambda_{k[12]}^{\ast}:=Cov\left(  F_{1}\left(  X^{(1)}\right)  ,P_{k-1}%
\left(  F_{2}\left(  X^{(2)}\right)  \right)  \right)  ,\text{ }k=2,3,...
\]
\noindent Similarly the $k$th \textit{L}-comoment of $F_{2}\left(  X^{\left(
2\right)  }\right)  $ with respect to $X^{\left(  1\right)  }$ is given by
\[
\lambda_{k[21]}^{\ast}=Cov\left(  F_{2}\left(  X^{(2)}\right)  ,P_{k-1}\left(
F_{1}\left(  X^{(1)}\right)  \right)  \right)  ,\text{ }k=2,3,...
\]
For the sake of notation simplicity, we set%
\begin{equation}
\delta_{k[12]}=\lambda_{\left(  k+1\right)  [12]}^{\ast}\text{ and }%
\delta_{k[21]}=\lambda_{\left(  k+1\right)  [21]}^{\ast},\text{ for }k=1,2,...
\label{deltak}%
\end{equation}
\noindent If the copula $C$ is symmetric in the sense that $C\left(
u,v\right)  =C\left(  v,u\right)  ,$ then\ $\delta_{k[12]}=\delta_{k[21]},$
for each $k=1,2...$ We call the coefficient $\delta_{k[12]}$ by "\textit{the
}$k$\textit{th bivariate copula L-moment}" of $X^{(1)}$ with respect to
$X^{(2)},$ similarly we call $\delta_{k[21]}$ by the $k$th copula
\textit{L}-moment of $X^{(2)}$ with respect to $X^{(1)}.$\medskip

\noindent In application, we will often make use of the three first bivariate
copula \textit{L}-moments, that is:\medskip%
\[%
\begin{tabular}
[c]{l}%
$\delta_{1[12]}=2Cov\left(  F_{1}\left(  X^{(1)}\right)  ,F_{2}\left(
X^{\left(  2\right)  }\right)  \right)  \bigskip$\\
$\delta_{2[12]}=-6Cov\left(  F_{1}\left(  X^{(1)}\right)  ,F_{2}\left(
X^{(2)}\right)  \left(  1-F_{2}\left(  X^{(2)}\right)  \right)  \right)
\bigskip$\\
$\delta_{3[12]}=Cov\left(  F_{1}\left(  X^{(1)}\right)  ,20F_{2}^{3}\left(
X^{(2)}\right)  -30F_{2}^{2}\left(  X^{(2)}\right)  +12F_{2}\left(
X^{(2)}\right)  -1\right)  .$%
\end{tabular}
\ \ \ \
\]

\section{\label{section3}\textbf{Bivariate copula representation of $k$th
copula \textit{L}-moment}}

\noindent Theorem \ref{TH1} below gives a representation of the $k$th
bivariate \textit{L}-moment in terms of the underlying copula function. This
result provides a new estimate of bivariate copula parameters.\medskip

\begin{theorem}
\label{TH1}\textit{The} $k$th bivariate copula \textit{\textit{L}-moment of
}$X^{(1)}$ with respect to $X^{(2)}$ \textit{may be rewritten, for each}
$k\geq1,$ \textit{as}%
\begin{equation}
\delta_{k\left[  12\right]  }=\int_{\mathbb{I}^{2}}\left(  C\left(
u_{1},u_{2}\right)  -u_{1}u_{2}\right)  du_{1}dP_{k}\left(  u_{2}\right)  ,
\label{eq4}%
\end{equation}
or%
\[
\delta_{k\left[  12\right]  }=\int_{\mathbb{I}^{2}}u_{1}P_{k}\left(
u_{2}\right)  dC\left(  u_{1},u_{2}\right)  .
\]

\end{theorem}

\noindent Observe that $\delta_{1\left[  12\right]  }=\delta_{1\left[
21\right]  }=\rho/6$ where $\rho$ is the Spearman rho $\rho,$ coefficient
defined in term of copula $C$ by
\begin{equation}
\rho=12\int_{\mathbb{I}^{2}}u_{1}u_{2}dC\left(  u_{1},u_{2}\right)  -3,
\label{rho}%
\end{equation}

\noindent(see Nelsen, 2006, page 167).

\noindent In view of Theorem \ref{TH1}, according to our needs, we can
construct a system of equations that will serve to the estimation of
multiparameter copula models. For this reason, the proposed estimator is more
likely to be used for the multiparameter copulas. In the case of the
one-parameter copulas, it is equivalent to the rho-inversion method (see
$(\ref{rho-inversion})$ below). Indeed, suppose that we are dealing with the
estimation of one dimension parameter of a copula model, then it suffices to
use one of the $k$th bivariate copula \textit{\textit{L}-}moment, says
$\delta_{1\left[  12\right]  }.\ $In the case of $r-$dimension parameters we
have to take the $r$ first bivariate copula \textit{\textit{L}-}moment, so we
obtain a system of $r$ equations with $r$ unknown parameters.\ Then, by
replacing the coefficients $\delta_{k\left[  12\right]  },$ $k=1,...,r$ by
their empirical counterparts, we obtain estimators of the $r$ parameters.
Indeed, suppose that $r=3$ and $C=C_{\theta},$ $\theta=\left(  \theta
_{1},\theta_{2},\theta_{3}\right)  ,$ then from Theorem \ref{TH1}, the first
three bivariate copula \textit{L}-moments of\textit{ }$X^{(1)}$ with respect
to $X^{(2)}$ are%

\begin{align*}
\delta_{1[12]}  &  =2\int_{\mathbb{I}^{2}}C_{\theta}\left(  u_{1}%
,u_{2}\right)  du_{1}du_{2}-\frac{1}{2}\medskip\\
\delta_{2\left[  12\right]  }  &  =6\int_{\mathbb{I}^{2}}\left(
2u_{2}-1\right)  C_{\theta}\left(  u_{1},u_{2}\right)  du_{1}du_{2}-\frac
{1}{2}\medskip\\
\delta_{3\left[  12\right]  }  &  =\int_{\mathbb{I}^{2}}\left(  60u_{2}%
^{2}-60u_{2}+12\right)  C_{\theta}\left(  u_{1},u_{2}\right)  du_{1}%
du_{2}-\frac{1}{2}.
\end{align*}

\noindent Since, in general the copula of $(X^{(1)},X^{(2)})$ is not the same
as of $(X^{(2)},X^{(1)})$, the corresponding parameters could be estimated
accordingly to ${\delta_{k[12]}}$ or ${\delta_{k[21]}}.$\medskip

\noindent Next we present applications of Theorem \ref{TH1} to parameter
estimation of two popular families of copula, namely the FGM and Archimedean copulas.

\subsection{FGM families}

\noindent One of the most popular parametric family of copulas is the FGM
family defined for $\left\vert \alpha\right\vert \leq1$ by%
\begin{equation}
C_{\alpha}\left(  u_{1},u_{2}\right)  =u_{1}u_{2}+\alpha u_{1}u_{2}%
\overline{u}_{1}\overline{u}_{2},\text{ }0\leq u_{1},u_{2}\leq1, \label{eq5}%
\end{equation}
with $\overline{u}_{j}:=1-u_{j},$ $j=1,2.$ The model is useful for the
moderate correlation which occurs in engineering and medical applications
(see, e.g., Blischke and Prabhaker Murthy, 2000 and Chalabian and Dunnington,
1998). The Pearson correlation coefficient $\rho$ corresponds to the model
$\left(  \ref{eq5}\right)  $ can never exceed $1/3,$ (see, e.g., Huang and
Kotz, 1984). In order to increase the dependence between two random variables
obeying the type of FGM distribution, Johnson and Kotz (1977) introduced the
$\left(  r-1\right)  $-iterated FGM family with $r$-dimensional parameter
$\mathbf{\alpha=}\left(  \alpha_{1},...,\alpha_{r}\right)  \mathbf{:}$%
\[
C_{\mathbf{\alpha}}\left(  u_{1},u_{2}\right)  =u_{1}u_{2}+\sum_{j=1}%
^{r}\alpha_{j}\left(  u_{1}u_{2}\right)  ^{\left[  j/2\right]  +1}\left(
\overline{u}_{1}\overline{u}_{2}\right)  ^{\left[  j/2+1/2\right]  },
\]
where $\left[  z\right]  $ denotes the greatest integer less than or equal to
$z.$ For example, the one-iterated FGM family (Huang and Kotz, 1984) is a
two-parameter copula model:%
\begin{equation}
C_{\alpha_{1},\alpha_{2}}\left(  u_{1},u_{2}\right)  =u_{1}u_{2}\left\{
1+\alpha_{1}\overline{u}_{1}\overline{u}_{2}+\alpha_{2}u_{1}u_{2}\overline
{u}_{1}\overline{u}_{2}\right\}  . \label{eq6}%
\end{equation}
\noindent The range of parameters $\left(  \alpha_{1},\alpha_{2}\right)  $ is
given by the region%
\begin{equation}
\mathcal{R}:=\left\{  \left(  \alpha_{1},\alpha_{2}\right)  ,\left\vert
\alpha_{1}\right\vert \leq1,\alpha_{1}+\alpha_{2}\geq-1,\text{ }\alpha_{2}%
\leq\frac{1}{2}\left[  3-\alpha_{1}+\left(  9-6\alpha_{1}-3\alpha_{1}%
^{2}\right)  ^{1/2}\right]  \right\}  . \label{eq7}%
\end{equation}
\noindent The maximal reached correlation for this family is
\begin{equation}
\rho_{FGM}^{\max}=0.42721,\text{ \ for }\left(  \alpha_{1},\alpha_{2}\right)
=\left(  -1+7/\sqrt{13},2-2/\sqrt{13}\right)  , \label{eq7*}%
\end{equation}
and the minimal correlation is $\rho_{FGM}^{\min}=-1/3$ for $\left(
\alpha_{1},\alpha_{2}\right)  =\left(  -1,0\right)  .$ The two-iterated FGM
family is given by%
\[
C_{\alpha_{1},\alpha_{2},\alpha_{3}}\left(  u_{1},u_{2}\right)  =u_{1}%
u_{2}\left\{  1+\alpha_{1}\overline{u}_{1}\overline{u}_{2}+\alpha_{2}%
u_{1}u_{2}\overline{u}_{1}\overline{u}_{2}+\alpha_{3}u_{1}u_{2}\left(
\overline{u}_{1}\overline{u}_{2}\right)  ^{2}\right\}  ,
\]
and it has been discussed by Lin (1987).\medskip

\noindent According to Theorem \ref{TH1}, we may give explicit formulas of
bivariate copula \textit{L}-moments for the FGM, the one-iterated FGM and the
two-iterated FGM.\ Since the number of parameters equals \textbf{ }%
$k\in\left\{  1,...,r\right\}  ,$ then we are dealing with first $k$ bivariate
copula \textit{L}-moments that will provide a system of $k$ equations and
therefore a tool for the estimation of the parameters of the copulas. We have :

\begin{itemize}
\item The first bivariate copula \textit{L}-moment of FGM family $C_{\alpha}$
is
\[
\delta_{1\left[  12\right]  }=\alpha/18
\]

\item The two first bivariate copula \textit{L}-moments of one-iterated FGM
copula $C_{\alpha_{1},\alpha_{2}}$ are:
\begin{equation}
\left\{
\begin{array}
[c]{l}%
\delta_{1\left[  12\right]  }=\alpha_{1}/18+\alpha_{2}/72\medskip\\
\delta_{2\left[  12\right]  }=\alpha_{2}/120
\end{array}
\right.  \label{eq8}%
\end{equation}

\item The three first bivariate copula \textit{L}-moments of two-iterated FGM
copula $C_{\alpha_{1},\alpha_{2},\alpha_{3}}$ are:%
\[
\left\{
\begin{array}
[c]{l}%
\delta_{1\left[  12\right]  }=\alpha_{1}/18+\alpha_{2}/72+\alpha
_{3}/450\medskip\\
\delta_{2\left[  12\right]  }=\alpha_{2}/120\medskip\\
\delta_{3\left[  12\right]  }=-\alpha_{3}/1050
\end{array}
\right.
\]

\end{itemize}

\subsection{Archimedean copula families}

\noindent The Archimedean copula family is one of important class of copula
models that contains the Gumbel, Clayton, Frank, ... (see, Table 4.1 in
Nelsen, 2006, page 116). In the bivariate case, an Archimedean copula is
defined by
\[
C(u,v)=\varphi^{-1}\left(  \varphi(u)+\varphi(v)\right)  ,
\]
where $\varphi:\mathbb{I}\rightarrow\mathbb{R}$ is a twice differentiable
function called the generator, satisfying: $\varphi\left(  1\right)  =0,$
$\varphi^{\prime}\left(  x\right)  <0,$ $\varphi^{\prime\prime}\left(
x\right)  \geq0$ for any $x\in\mathbb{I}/\left\{  0,1\right\}  .$ The notation
$\varphi^{-1}$ stands for the inverse function of $\varphi.$ For examples, the
three generators $\varphi_{\theta}\left(  t\right)  =\left(  -\ln\left(
\left(  1-\theta\left(  1-t\right)  \right)  /t\right)  \right)  ,$
$\varphi_{\alpha}\left(  t\right)  =\left(  t^{-\alpha}-1\right)  /\alpha$ and
$\varphi_{\beta}\left(  t\right)  =\left(  -\ln t\right)  ^{\beta}$ define,
respectively, the one parameter Frank, Clayton and Gumbel copula families. For
example, the Gumbel family is defined by%
\begin{equation}
C_{\beta}(u,v)=\exp\left(  -\left[  \left(  -\ln u\right)  ^{\beta}+\left(
-\ln v\right)  ^{\beta}\right]  ^{1/\beta}\right)  ,\text{ }\beta\geq1.
\label{gumbel1}%
\end{equation}
\noindent For more flexibility in fitting data, it is better to use the
multi-parameters copula models than those of one parameter. To have a copula
with more one parameter, we use, for instance, the distorted copula defined by
$C_{\Gamma}\left(  u,v\right)  =\Gamma^{-1}\left(  C\left(  \Gamma\left(
u\right)  ,\Gamma\left(  v\right)  \right)  \right)  ,$ where $\Gamma
:\mathbb{I}\rightarrow\mathbb{I}$ is a continuous, concave and strictly
increasing function with $\Gamma\left(  0\right)  =0$ and $\Gamma\left(
1\right)  =1.$ Note that if $C$ is an Archimedean copula with generator
$\varphi,$ then $C_{\Gamma}$ is also Archimedean copula with generator
generator $\varphi\circ\Gamma.\ $For more details see Nelsen (2006), page 96 .
As example, suppose that $\Gamma=\Gamma_{\beta_{2}},$ with $\Gamma_{\beta_{2}%
}\left(  t\right)  =\exp\left(  t^{-\beta_{2}}-1\right)  ,$ $\beta_{2}>0$ and
consider a Gumbel copula $C_{\beta_{1}}$ with generator $\varphi_{\beta
1}\left(  t\right)  =\left(  -\ln t\right)  ^{\beta_{1}},$ $\beta_{1}\geq1.$
Then the copula $C_{\beta_{1},\beta_{2}}\left(  u,v\right)  =\Gamma_{\beta
_{2}}^{-1}\left(  C_{\beta_{1}}\left(  \Gamma_{\beta_{2}}\left(  u\right)
,\Gamma_{\beta_{2}}\left(  v\right)  \right)  \right)  $ given by%
\begin{equation}
C_{\beta_{1},\beta_{2}}\left(  u,v\right)  :=\left(  \left(  \left(
u^{-\beta_{2}}-1\right)  ^{\beta_{1}}+\left(  v^{-\beta_{2}}-1\right)
^{\beta_{1}}\right)  ^{1/\beta_{1}}+1\right)  ^{1/\beta_{2}}, \label{gumbel}%
\end{equation}
which is a two-parameter Archimedean copula with generator $\varphi_{\beta
_{1},\beta_{2}}\left(  t\right)  :=\left(  t^{-\beta_{2}}-1\right)
^{\beta_{1}}.$ (see Nelsen 2006, page 96).\medskip

\noindent To have the two first bivariate copula \textit{L}-moments correspond
to $C_{\beta_{1},\beta_{2}},$ we apply Theorem \ref{TH1} to get the following
system of equations:%
\begin{equation}
\left\{
\begin{array}
[c]{l}%
\delta_{1\left[  12\right]  }=2\int_{0}^{1}\int_{0}^{1}\left(  C_{\beta
_{1},\beta_{2}}\left(  u,v\right)  -uv\right)  dudv,\medskip\\
\delta_{2\left[  12\right]  }=6\int_{0}^{1}\int_{0}^{1}\left(  2v-1\right)
\left(  C_{\beta_{1},\beta_{2}}\left(  u,v\right)  -uv\right)  dudv.
\end{array}
\right.  \label{gumbelsystem}%
\end{equation}
\noindent In this case we cannot give explicit formulas, in terms of $\left\{
\delta_{1\left[  12\right]  },\delta_{2\left[  12\right]  }\right\}  ,$ for
the parameters $\left\{  \beta_{1},\beta_{2}\right\}  ,$ however for given
values of the\ bivariate copula \textit{L}-moments, we can obtain the
corresponding values of $\left\{  \beta_{1},\beta_{2}\right\}  $ by solving
the previous system by numerical methods.

\begin{remark}
The previous system provides estimators for copula parameters by replacing the
bivariate copula \textit{L}-moments by their sample counterparts. This is
similar to the method of moments (see Section \ref{section4}).
\end{remark}

\section{\label{section4}\textbf{Semi-parametric BLM-based estimation}}

\noindent The aim of the present section is to provide a semi-parametric
estimation for bivariate copula parameters on the basis of results of Section
\ref{section3}. Suppose that the underlying copula $C$ belongs to a parametric
family $C_{\theta}$ with $\mathbf{\theta}=(\theta_{1},\cdots,\theta_{r}%
)\in\mathcal{O}$ an open subset of $\mathbb{R}^{r}$ satisfies the
\textit{concordance ordering condition }of copulas\textit{ }(see, Nelsen,
2006, page 135), that is:%
\begin{equation}
\text{for every }\mathbf{\theta}^{\left(  1\right)  },\mathbf{\theta}^{\left(
2\right)  }\in\mathcal{O}:\text{ }\mathbf{\theta}^{\left(  1\right)  }%
\neq\mathbf{\theta}^{\left(  2\right)  }\Longrightarrow C_{\mathbf{\theta
}^{\left(  1\right)  }}\left(  >\text{ or }<\right)  C_{\mathbf{\theta
}^{\left(  2\right)  }}. \label{order}%
\end{equation}
\noindent The above inequalities mean that for any $\left(  u,v\right)
\in\mathbb{I}^{2},$ $C_{\mathbf{\theta}^{\left(  1\right)  }}\left(
u,v\right)  \left(  >\text{ or }<\right)  C_{\mathbf{\theta}^{\left(
2\right)  }}\left(  u,v\right)  .$\ It is clear that condition (\ref{order})
implies the well known \textit{identifiability condition} of copulas:%
\begin{equation}
\text{for every }\mathbf{\theta}^{\left(  1\right)  },\mathbf{\theta}^{\left(
2\right)  }\in\mathcal{O}:\text{ }\mathbf{\theta}^{\left(  1\right)  }%
\neq\mathbf{\theta}^{\left(  2\right)  }\Longrightarrow C_{\mathbf{\theta
}^{\left(  1\right)  }}\neq C_{\mathbf{\theta}^{\left(  2\right)  }}.
\label{identif}%
\end{equation}
\noindent Identifiability is a natural and even a necessary condition: if the
parameter is note identifiable then consistent estimator cannot exist (see,
e.g., van der Vaart, 1998, page 62). A wide class of copula families satisfy
condition (\ref{order}), and therefore (\ref{identif}), including the iterated
FGM and the archimedean models. Indeed, the iterated FGM family is linear with
respect to their parameters, then by a little algebra we easily verify this
condition. For two-parameter Gumbel family,\ the condition (\ref{order}) is
already checked in Nelsen (2006), page 144, example 4.22. Consider now a
random sample $\left(  X_{i}^{(1)},X_{i}^{(2)}\right)  _{i=1,n}$ from the
bivariate r.v. $\left(  X^{(1)},X^{(2)}\right)  .$ For each $j=1,2,$ let
$F_{j:n}^{+}:=nF_{j:n}/\left(  n+1\right)  $ denotes the rescaled empirical df
corresponds to the empirical df%
\[
F_{n:j}\left(  x_{j}\right)  =n^{-1}\sum_{i=1}^{n}\mathbf{1}\left\{
X_{i}^{(j)}\leq x_{j}\right\}  .
\]

\noindent We are now in position to present, in three steps, the
semi-parametric BLM-based estimation:

\begin{itemize}
\item \textit{Step} 1: For each $k=1,...,r,$ compute%
\begin{equation}
\widehat{\delta}_{k[12]}=n^{-1}\sum_{i=1}^{n}F_{1:n}^{+}\left(  X_{i}%
^{(1)}\right)  P_{k}\left(  F_{2:n}^{+}\left(  X_{i}^{(2)}\right)  \right)  .
\label{gamacha}%
\end{equation}
given in equation (\ref{deltak}).

\item \textit{Step 2}: Using Theorem \ref{TH1} to generate a system of $r$
equations given by equation (\ref{eq4}), for $k=1,...,r.$

\item \textit{Step 3:} Solve the system%
\begin{equation}
\left\{
\begin{array}
[c]{l}%
\delta_{1[12]}\left(  \theta_{1},...,\theta_{r}\right)  =\widehat{\delta
}_{1[12]}\\
\delta_{2[12]}\left(  \theta_{1},...,\theta_{r}\right)  =\widehat{\delta
}_{2[12]}\\
\vdots\\
\delta_{r[12]}\left(  \theta_{1},...,\theta_{r}\right)  =\widehat{\delta
}_{r[12]}.
\end{array}
\ \right.  \label{system1}%
\end{equation}

\noindent The obtained solution $\mathbf{\hat{\theta}}^{BLM}:=\left(
\hat{\theta}_{1},...,\hat{\theta}_{r}\right)  $ is called a BLM estimator for
$\mathbf{\theta}=\left(  \theta_{1},...,\theta_{r}\right)  .\medskip$
\end{itemize}

\noindent The existence and the convergence of a solution of the previous
system are established in Theorem \ref{TH2}, (see Section \ref{section4.2}).
$\medskip$

\noindent As an application of the BLM based estimation, we choose the
one-iterated FGM copula $C_{\alpha_{1},\alpha_{2}}$ given in (\ref{eq6}) and
propose estimators for the parameters $\left(  \alpha_{1},\alpha_{2}\right)  $
noted $\left(  \widehat{\alpha}_{1},\widehat{\alpha}_{2}\right)  .$ For this
family, using (\ref{eq8})\textbf{, }system (\ref{system1}) becomes%
\[
\left\{
\begin{array}
[c]{l}%
\alpha_{1}/18+\alpha_{2}/72=\widehat{\delta}_{1\left[  12\right]  },\medskip\\
\alpha_{2}/120=\widehat{\delta}_{2\left[  12\right]  },\medskip
\end{array}
\right.
\]
where $\widehat{\delta}_{k[12]},$ $k=1,2$ are given in (\ref{gamacha}).
Therefore
\[
\left\{
\begin{array}
[c]{l}%
\widehat{\alpha}_{1}=18\widehat{\delta}_{1\left[  12\right]  }-30\widehat
{\delta}_{2\left[  12\right]  },\medskip\\
\widehat{\alpha}_{2}=120\widehat{\delta}_{2\left[  12\right]  }.
\end{array}
\right.
\]

\subsection{BLM as a rank approximate \textit{Z}-estimation}

\noindent Tsukahara (2005) introduced a new estimation method for copula
models called the rank approximate $Z$-estimation (RAZ) that generalizes the
PML one.\ The BLM method may be interpreted as a RAZ estimation. Let
$\Psi\left(  \cdot;\mathbf{\theta}\right)  $ be an $\mathbb{R}^{r}$-valued
function on $\mathbb{I}^{2},$ called "score function", whose components
$\Psi_{j}\left(  \cdot;\mathbf{\theta}\right)  $ satisfy the condition%
\[
\int_{\mathbb{I}^{2}}\Psi_{j}\left(  u_{1},u_{2};\mathbf{\theta}\right)
dC\left(  u_{1},u_{2}\right)  =0,\text{ for }j=1,...,r.
\]
Any solution $\widehat{\mathbf{\theta}}^{RAZ}$ in $\mathbf{\theta}$ of the
following equation%
\begin{equation}
\sum_{i=1}^{n}\Psi\left(  F_{1:n}^{+}\left(  X_{i}^{\left(  1\right)
}\right)  ,F_{2:n}^{+}\left(  X_{i}^{\left(  2\right)  }\right)
;\mathbf{\theta}\right)  =0, \label{eq9}%
\end{equation}
is called a RAZ estimator. There may not be an exact solution to equation
$(\ref{eq9})$ in general, so in practice, we should choose $\widehat
{\mathbf{\theta}}^{RAZ}$ to be any value of $\theta$ for which the absolute
value of the left-hand side of equation $(\ref{eq9})$ is close to zero. It is
worth mentioning that if the copula $C_{\mathbf{\theta}}$ is absolutely
continuous with density $c_{\mathbf{\theta}},$ then the function
$\Psi=\overset{\bullet}{c}_{\mathbf{\theta}}/c_{\mathbf{\theta}},$ with
$\overset{\bullet}{c}_{\mathbf{\theta}}=\left(  \partial c_{\mathbf{\theta}%
}/\partial\theta_{k}\right)  _{j=1,...,r},$ leads the PML based estimation
(see for instance Genest \textit{et al}., 1995). The existence of a sequence
of consistent roots of $Z$-estimation in this context is discussed in Theorem
1 in Tsukahara (2005). One of the main question of RAZ-estimation is the
choice of the score function $\Psi$ producing, in a certain sense, the best
estimator.\ In Section (\ref{simulation}), we show that the the copula
\textit{L}-moment score (CLS) functions (\ref{Lk})\textit{ }improve the
concordance score functions in terms of bias and root mean square error
(RMSE).\medskip

\noindent The univariate L-moments are a special case of the L-statistics with
a specific choice of the weight coefficients. This fact makes the L-moments
interpretable and popular. In a similar way, the BLM-based estimator
represents an important special case of the RAZ one. In our case, the score
function $\Psi\left(  u_{1},u_{2};\mathbf{\theta}\right)  $ corresponds to the
function $L_{k}$ given in (\ref{Lk}) below. Recall that from Theorem \ref{TH1}
we have%
\[
\delta_{k\left[  12\right]  }\left(  \mathbf{\theta}\right)  =\int
_{\mathbb{I}^{2}}u_{1}P_{k}\left(  u_{2}\right)  dC_{\mathbf{\theta}}\left(
u_{1},u_{2}\right)  ,\text{ }k=1,2...,
\]
and define the CLS\textit{ }functions by
\begin{equation}
L_{k}\left(  u_{1},u_{2};\mathbf{\theta}\right)  :=u_{1}P_{k}\left(
u_{2}\right)  -\delta_{k\left[  12\right]  }\left(  \mathbf{\theta}\right)
,\text{ }k=1,...,r, \label{Lk}%
\end{equation}
satisfying $\int_{\mathbb{I}^{2}}L_{k}\left(  u_{1},u_{2};\mathbf{\theta
}\right)  dC_{\mathbf{\theta}}\left(  u_{1},u_{2}\right)  =0,$ $k=1,...,r.$
Then the RAZ estimator corresponding to the CLS function $\mathbf{L=}\left(
L_{1},...,L_{r}\right)  $ is a solution in $\mathbf{\theta}$\ of the system%
\begin{equation}
\sum_{i=1}^{n}\mathbf{L}\left(  F_{1:n}^{+}\left(  X_{i}^{\left(  1\right)
}\right)  ,F_{2:n}^{+}\left(  X_{i}^{\left(  2\right)  }\right)
;\mathbf{\theta}\right)  =0, \label{LK*}%
\end{equation}
that is%
\begin{equation}
\sum_{i=1}^{n}F_{1:n}^{+}\left(  X_{i}^{(1)}\right)  P_{k}\left(  F_{2:n}%
^{+}\left(  X_{i}^{(2)}\right)  \right)  -n\delta_{k\left[  12\right]
}\left(  \mathbf{\theta}\right)  =0,\text{ }k=1,...,r, \label{eq11}%
\end{equation}
therefore $\delta_{k[12]}\left(  \mathbf{\theta}\right)  =\widehat{\delta
}_{k[12]},$ $k=1,...,r,$ which in fact the system of bivariate copula
\textit{L}-moments given in system (\ref{system1}).\medskip

\noindent On the other hand, the measures of concordance produce also a
$Z$-estimation for copula models. Indeed, the most popular measures of
concordance (see, Nelsen, 2006, page 182) are Kendall's tau $(\tau),$
Spearman's rho $\left(  \rho\right)  ,$ Gini's gamma $\left(  \gamma\right)  $
and Spearman's foot-rule phi $\left(  \varphi\right)  ,$ given respectively by%

\begin{align*}
\tau\left(  \mathbf{\theta}\right)   &  =4\int_{\mathbb{I}^{2}}C_{\theta
}\left(  u_{1},u_{2}\right)  dC_{\mathbf{\theta}}\left(  u_{1},u_{2}\right)
-1,\medskip\\
\rho\left(  \mathbf{\theta}\right)   &  =12\int_{\mathbb{I}^{2}}u_{1}%
u_{2}dC_{\mathbf{\theta}}\left(  u_{1},u_{2}\right)  -3,\medskip\\
\gamma\left(  \mathbf{\theta}\right)   &  =4\int_{\mathbb{I}}C_{\mathbf{\theta
}}\left(  u_{1},1-u_{1}\right)  du_{1}-\int_{\mathbb{I}}\left(  u_{1}%
-C_{\mathbf{\theta}}\left(  u_{1},u_{1}\right)  \right)  du_{1},\medskip\\
\varphi\left(  \mathbf{\theta}\right)   &  =1-3\int_{\mathbb{I}^{2}}\left\vert
u_{1}-u_{2}\right\vert dC_{\mathbf{\theta}}\left(  u_{1},u_{2}\right)  .
\end{align*}

\noindent It follows that the concordance score (CS) functions associated to
$\tau,$ $\rho,$ $\gamma$ and $\varphi$ respectively are%

\begin{align*}
\Psi_{1}\left(  u_{1},u_{2};\mathbf{\theta}\right)   &  :=4C_{\mathbf{\theta}%
}\left(  u_{1},u_{2}\right)  -\tau\left(  \mathbf{\theta}\right)  ,\bigskip\\
\Psi_{2}\left(  u_{1},u_{2};\mathbf{\theta}\right)   &  :=12u_{1}u_{2}%
-3-\rho\left(  \theta\right)  ,\bigskip\\
\Psi_{3}\left(  u_{1},u_{2};\mathbf{\theta}\right)   &  :=4C_{\mathbf{\theta}%
}\left(  u_{1},1-u_{1}\right)  -u_{1}+C_{\mathbf{\theta}}\left(  u_{1}%
,u_{1}\right)  -\gamma\left(  \mathbf{\theta}\right)  ,\bigskip\\
\Psi_{4}\left(  u_{1},u_{2};\mathbf{\theta}\right)   &  :=1-3\left\vert
u_{1}-u_{2}\right\vert -\varphi\left(  \mathbf{\theta}\right)  .
\end{align*}
It is clear that $\int_{\mathbb{I}^{2}}\Psi_{j}\left(  u_{1},u_{2}%
;\mathbf{\theta}\right)  dC_{\mathbf{\theta}}\left(  u_{1},u_{2}\right)  =0,$
$j=1,...,4,$\ then whenever the dimension of parameters $r=4,$ the function
$\Psi=\left(  \Psi_{1},...,\Psi_{4}\right)  $ provides $Z$-estimators for
copula models. If the dimension of parameters $r<4,$ then we may choose any
$r$ functions from $\Psi_{1},...,\Psi_{4}$ to have a system of $r$
equations\textbf{ }that provides estimators of the\textbf{ }$r$\textbf{
}parameters. $\medskip$

\noindent Tsukahara (2005) also discussed the RAZ-estimators based on
Kendall's tau $(\tau)$ and Spearman's rho $\left(  \rho\right)  ,$ called
$\tau$-score and $\rho$-score RAZ-estimators.\ Suppose that $r=1$ and let
$\widehat{\tau}_{n}$ and $\widehat{\rho}_{n}$ be the sample versions of
Kendall's tau $(\tau)$ and Spearman's rho $\left(  \rho\right)  .$\ By using
the same idea as the method of moments, the $\tau$-inversion $\widehat{\theta
}_{\tau}$ estimator and the $\rho$-inversion $\widehat{\theta}_{\rho}$
estimator of $\theta$ are defined by%
\begin{equation}
\widehat{\theta}_{\tau}=\tau^{-1}\left(  \widehat{\tau}_{n}\right)  \text{ and
}\widehat{\theta}_{\rho}=\rho^{-1}\left(  \widehat{\rho}_{n}\right)  .
\label{rho-inversion}%
\end{equation}
In the case when $r=2,$ we may also estimate $\mathbf{\theta=}\left(
\theta_{1},\theta_{2}\right)  $ by solving the system%
\begin{equation}
\left\{
\begin{array}
[c]{l}%
\tau\left(  \theta_{1},\theta_{2}\right)  =\widehat{\tau}_{n}\medskip\\
\rho\left(  \theta_{1},\theta_{2}\right)  =\widehat{\rho}_{n}.
\end{array}
\right.  \label{rho-tau-sys}%
\end{equation}
We call, the obtaining solution of the previous system, by $\left(  \tau
,\rho\right)  -$inversion estimator of $\mathbf{\theta}.$ Suppose that we are
dealing with the estimation of parameters $\left(  \alpha_{1},\alpha
_{2}\right)  $ of the one-iterated FGM copula $C_{\alpha_{1},\alpha_{2}}$ in
$(\ref{eq6}).\ $Then, the associated Kendall's tau $(\tau)$ and Spearman's rho
$\left(  \rho\right)  $ are%
\begin{equation}
\left\{
\begin{tabular}
[c]{l}%
$\tau\left(  \alpha_{1},\alpha_{2}\right)  =2\alpha_{1}/9+\alpha_{2}%
/18+\alpha_{1}\alpha_{2}/450\medskip$\\
$\rho\left(  \alpha_{1},\alpha_{2}\right)  =\alpha_{1}/3+\alpha_{2}/12.$%
\end{tabular}
\ \ \ \ \ \ \ \right.  \label{eq10}%
\end{equation}
The $\left(  \tau,\rho\right)  $-inversion estimator of parameters $\left(
\alpha_{1},\alpha_{2}\right)  $ is the solution of the system%
\[
\left\{
\begin{array}
[c]{l}%
\tau\left(  \widehat{\alpha}_{1},\widehat{\alpha}_{2}\right)  =\widehat{\tau
}_{n}\medskip\\
\rho\left(  \widehat{\alpha}_{1},\widehat{\alpha}_{2}\right)  =\widehat{\rho
}_{n}.
\end{array}
\right.
\]
Similarly, if we consider the FGM family $C_{\alpha_{1},\alpha_{2},\alpha_{3}%
}$ we have to add $\gamma-$score and $\varphi-$score to have a system of four
equations, we omit details.\medskip

\subsection{\label{section4.2}Asymptotic behavior of the BLM estimator}

\noindent By considering BLM's estimator as a RAZ-estimator, a straight
application of Theorem 1 in Tsukahara (2005) leads to the consistency and
asymptotic normality of the considered estimator. Let $\mathbf{\theta}_{0}$ be
the true value of $\mathbf{\theta}$ and assume that the assumptions $\left[
H.1\right]  -\left[  H.3\right]  $ listed below are fulfilled.$\medskip$

\begin{itemize}
\item $\left[  H.1\right]  $ $\mathbf{\theta}_{0}\in\mathcal{O}\subset
\mathbb{R}^{r}$ is the unique zero of the mapping $\mathbf{\theta}%
\rightarrow\int_{\mathbb{I}^{2}}\mathbf{L}\left(  u_{1},u_{2};\mathbf{\theta
}\right)  dC_{\mathbf{\theta}_{0}}\left(  u_{1},u_{2}\right)  $ that is
defined from $\mathcal{O}$ to $\mathbb{R}^{r}.$

\item $\left[  H.2\right]  $ $\mathbf{L}\left(  \cdot;\mathbf{\theta}\right)
$ is differentiable with respect to $\mathbf{\theta}$ with the Jacobian matrix
denoted by%
\[
\overset{\bullet}{\mathbf{L}}\left(  u_{1},u_{2};\mathbf{\theta}\right)
:=\left[  \frac{\partial L_{k}\left(  u_{1},u_{2};\mathbf{\theta}\right)
}{\partial\theta_{\ell}}\right]  _{r\times r},
\]
$\overset{\bullet}{\mathbf{L}}\left(  u_{1},u_{2};\mathbf{\theta}\right)  $ is
continuous both in $\left(  u_{1},u_{2}\right)  $ and $\mathbf{\theta},$ and
the Euclidian norm $\left\vert \overset{\bullet}{\mathbf{L}}\left(
u_{1},u_{2};\mathbf{\theta}\right)  \right\vert $ is dominated by a
$dC_{\mathbf{\theta}}$-integrable function $h\left(  u_{1},u_{2}\right)  .$

\item $\left[  H.3\right]  $ The $r\times r$ matrix $A_{0}:=\int
_{\mathbb{I}^{2}}\overset{\bullet}{\mathbf{L}}\left(  u_{1},u_{2}%
;\mathbf{\theta}_{0}\right)  dC_{\mathbf{\theta}_{0}}\left(  u_{1}%
,u_{2}\right)  $ is nonsingular.$\medskip$
\end{itemize}

\begin{theorem}
\label{TH2}Assume that the assumptions $\left[  H.1\right]  -\left[
H.3\right]  $ hold. Then with probability tending to one as $n\rightarrow
\infty,$ there exists a solution $\widehat{\mathbf{\theta}}^{BLM}$ to the
equation $\left(  \ref{eq11}\right)  $ which converges to $\mathbf{\theta}%
_{0}.$ Moreover
\[
\sqrt{n}\left(  \widehat{\mathbf{\theta}}^{BLM}-\mathbf{\theta}_{0}\right)
\overset{\mathcal{D}}{\rightarrow}\mathcal{N}\left(  \mathbf{0}_{\mathbb{R}%
^{r}},A_{0}^{-1}%
{\textstyle\sum\nolimits_{0}}
\left(  A_{0}^{-1}\right)  ^{T}\right)  ,\text{ as }n\rightarrow\infty,
\]
with
\begin{equation}%
{\textstyle\sum\nolimits_{0}}
:=var\left\{  \mathbf{L}\left(  \xi_{1},\xi_{2};\mathbf{\theta}_{0}\right)
+\sum_{j=1}^{2}\int_{\mathbb{I}^{2}}\mathbf{M}_{j}\left(  u_{1},u_{2}\right)
\left(  \mathbf{1}\left\{  \xi_{j}\leq u_{j}\right\}  -u_{j}\right)
dC_{\mathbf{\theta}_{0}}\left(  u_{1},u_{2}\right)  \right\}  , \label{sigma}%
\end{equation}
where $\left(  \xi_{1},\xi_{2}\right)  $ is a bivariate r.v. with joint
distribution function $C_{\theta_{0}},$%
\[
\mathbf{M}_{1}\left(  u_{1},u_{2}\right)  :=\left\{  P_{k}\left(
u_{2}\right)  \right\}  _{k=1,r}\text{ and }\mathbf{M}_{2}\left(  u_{1}%
,u_{2}\right)  :=\left\{  u_{1}dP_{k}\left(  u_{2}\right)  /du_{2}\right\}
_{k=1,r}.
\]

\end{theorem}

\noindent The proof of Theorem \ref{TH2} is given in the Appendix. \medskip

\noindent Note that the assumptions $\left[  H.1\right]  -\left[  H.3\right]
$ correspond to $A2,$ $A4$ and $A5$ in Tsukahara (2005). The score function
$\mathbf{L}$ meets the remaining assumptions $A1$ and $A3.$

\subsection{A discussion on Theorem 4.1}

\noindent Notice that assumption $\left[  H.1\right]  $ is verified for any
parametric copula $C_{\mathbf{\theta}}$ satisfying the concordance ordering
condition of copulas $\left(  \ref{order}\right)  .$ Indeed, suppose that
there exists $\mathbf{\theta}_{1}\neq\mathbf{\theta}_{0},$ such that
\begin{equation}
\int_{\mathbb{I}^{2}}L_{k}\left(  u_{1},u_{2};\mathbf{\theta}_{1}\right)
dC_{\mathbf{\theta}_{0}}\left(  u_{1},u_{2}\right)  =0,\text{ for every }%
k\in\left\{  1,...,r\right\}  . \label{LK}%
\end{equation}

\noindent Recall that $\delta_{k[12]}\left(  \mathbf{\theta}_{0}\right)
=\int_{\mathbb{I}^{2}}u_{1}P_{k}\left(  u_{2}\right)  dC_{\mathbf{\theta}_{0}%
}\left(  u_{1},u_{2}\right)  $ and $\delta_{k[12]}\left(  \mathbf{\theta}%
_{1}\right)  =\int_{\mathbb{I}^{2}}u_{1}P_{k}\left(  u_{2}\right)
dC_{\mathbf{\theta}_{1}}\left(  u_{1},u_{2}\right)  $ and, from assumption
$\left(  \ref{order}\right)  ,$ $C_{\mathbf{\theta}_{0}}\left(  >\text{ or
}<\right)  C_{\mathbf{\theta}_{1}}.$ It follows, by monotonicity of the
integral, that $\delta_{k[12]}\left(  \mathbf{\theta}_{1}\right)  \left(
>\text{or}<\right)  \delta_{k[12]}\left(  \mathbf{\theta}_{0}\right)  ,$ this
implies that $\delta_{k[12]}\left(  \mathbf{\theta}_{1}\right)  -\delta
_{k[12]}\left(  \mathbf{\theta}_{0}\right)  \neq0,$ for every $k\in\left\{
1,...,r\right\}  .$ Observe that $\int_{\mathbb{I}^{2}}dC_{\mathbf{\theta}%
_{0}}\left(  u_{1},u_{2}\right)  =1$ then $\delta_{k[12]}\left(
\mathbf{\theta}_{1}\right)  =\int_{\mathbb{I}^{2}}\delta_{k[12]}\left(
\mathbf{\theta}_{1}\right)  dC_{\mathbf{\theta}_{0}}\left(  u_{1}%
,u_{2}\right)  ,$ consequently, since $L_{k}\left(  u_{1},u_{2};\mathbf{\theta
}_{1}\right)  =u_{1}P_{k}\left(  u_{2}\right)  -\delta_{k[12]}\left(
\mathbf{\theta}_{1}\right)  ,$ that%

\begin{align*}
\int_{\mathbb{I}^{2}}L_{k}\left(  u_{1},u_{2};\mathbf{\theta}_{1}\right)
dC_{\mathbf{\theta}_{0}}\left(  u_{1},u_{2}\right)   &  =\int_{\mathbb{I}^{2}%
}u_{1}P_{k}\left(  u_{2}\right)  dC_{\mathbf{\theta}_{0}}\left(  u_{1}%
,u_{2}\right)  -\int_{\mathbb{I}^{2}}\delta_{k[12]}\left(  \mathbf{\theta}%
_{1}\right)  dC_{\mathbf{\theta}_{0}}\left(  u_{1},u_{2}\right) \\
&  =\delta_{k[12]}\left(  \mathbf{\theta}_{0}\right)  -\delta_{k[12]}\left(
\mathbf{\theta}_{1}\right)  \neq0,\text{ for every }k\in\left\{
1,...,r\right\}  ,
\end{align*}
which is a contradiction with equation $\left(  \ref{LK}\right)  ,$ as sought.
Let's now discuss the rest of assumptions.$\ $In $\left[  H.2\right]  ,$ the
continuity and the differentiability with respect to\ $\mathbf{\theta}$ and
$\left(  u_{1},u_{2}\right)  $ of $\mathbf{L}\left(  \cdot;\mathbf{\theta
}\right)  $ and $\overset{\bullet}{\mathbf{L}}\left(  \cdot;\mathbf{\theta
}\right)  $ are lie with that of copula $C_{\mathbf{\theta}},$ which are
natural assumptions in parametric copula models. Some examples on this issue
are illustrated in Fredricks \textit{et al.} (2007). The second part of
$\left[  H.2\right]  $ and $\left[  H.3\right]  $ may be checked for a given
copula model. For example, if we consider the FGM family (see $\left(
\ref{eq5}\right)  $) we get $L_{1}\left(  u_{1},u_{2};\alpha\right)
=u_{1}\left(  2u_{2}-1\right)  -\alpha/18$ and $L_{k}\left(  u_{1}%
,u_{2};\alpha\right)  =0,$ for $k=2,3...$Then $dL_{1}\left(  u_{1}%
,u_{2};\alpha\right)  /d\alpha=-1/18$ and $dL_{k}\left(  u_{1},u_{2}%
;\alpha\right)  /d\alpha=0,$ for $k=2,...$\ Let $\alpha_{0}$ denote the true
value of parameter $\alpha.$ It is clear that each compound of $\overset
{\bullet}{\mathbf{L}}$ is continuous with respect to $\alpha$ and $\left(
u_{1},u_{2}\right)  ,$ $\left\vert \overset{\bullet}{\mathbf{L}}\left(
u_{1},u_{2};\alpha\right)  \right\vert =1/18,$ which is $C_{\alpha}%
-$integrable function and, $A_{0}=\int_{\mathbb{I}^{2}}\overset{\bullet
}{\mathbf{L}}\left(  u_{1},u_{2};\alpha_{0}\right)  dC_{\alpha_{0}}=-1/18$
which is nonsingular matrix, then the assumptions $\left[  H.2\right]  $ and
$\left[  H.3\right]  $ are well verified. By a little algebra we get to the
corresponding value of $%
{\textstyle\sum\nolimits_{0}}
$ that is defined in $\left(  \ref{sigma}\right)  ,$ and by Theorem \ref{TH2}
we get
\[
\sqrt{n}\left(  \widehat{\alpha}^{BLM}-\alpha_{0}\right)  \overset
{\mathcal{D}}{\rightarrow}\mathcal{N}\left(  0,\alpha_{0}^{2}/270+1/5\right)
,\text{ as }n\rightarrow\infty.
\]
For the one-iterated FGM family (see $\left(  \ref{eq6}\right)  $), by letting
$\alpha_{1}=\alpha$ and $\alpha_{2}=\beta,$ it is readily to verify that%

\begin{align*}
L_{1}\left(  u_{1},u_{2};\alpha,\beta\right)   &  =u_{1}\left(  2u_{2}%
-1\right)  -\alpha/18-\beta/72,\medskip\\
L_{2}\left(  u_{1},u_{2};\alpha,\beta\right)   &  =u_{1}\left(  6u_{2}%
^{2}-6u_{2}+1\right)  -\beta/120,
\end{align*}

\noindent which, obviously, are continuous with respect to $\left(
\alpha,\beta\right)  $ and $\left(  u_{1},u_{2}\right)  $ and $C_{\alpha
,\beta}-$integrable function, and
\[
\overset{\bullet}{\mathbf{L}}\left(  u_{1},u_{2};\alpha,\beta\right)  =\left[
\begin{array}
[c]{cc}%
-1/18 & -1/72\\
0 & -1/120
\end{array}
\right]  .
\]
Let $\left(  \alpha_{0},\beta_{0}\right)  $ denote the true value of parameter
$\left(  \alpha,\beta\right)  $ and by calculating the elements of the matrix
\[
A_{0}:=\int_{\mathbb{I}^{2}}\overset{\bullet}{\mathbf{L}}\left(  u_{1}%
,u_{2};\alpha_{0},\beta_{0}\right)  dC_{\alpha_{0},\beta_{0}},
\]
we get%
\[
A_{0}=\left[
\begin{array}
[c]{cc}%
-1/18 & 1/72\\
0 & -1/120
\end{array}
\right]  ,
\]
which is nonsingular because its determinant equals $1/2160\neq0,\ $therefore
$\left[  H.2\right]  $ and $\left[  H.3\right]  $ are also\ verified. Then, in
view of Theorem \ref{TH2}, we have
\[
\sqrt{n}\left\{  \left(
\begin{array}
[c]{c}%
\widehat{\alpha}^{BLM}\\
\widehat{\beta}^{BLM}%
\end{array}
\right)  -\left(
\begin{array}
[c]{c}%
\alpha_{0}\\
\beta_{0}%
\end{array}
\right)  \right\}  \overset{\mathcal{D}}{\rightarrow}\mathcal{N}\left(
\left(
\begin{array}
[c]{c}%
0\\
0
\end{array}
\right)  ,%
{\textstyle\sum\nolimits^{2}}
\right)  ,\text{ as }n\rightarrow\infty,
\]
where $%
{\textstyle\sum\nolimits^{2}}
:=A_{0}^{-1}%
{\textstyle\sum\nolimits_{0}}
\left(  A_{0}^{-1}\right)  ^{T}.\ $After a tedious computation we
get$\medskip\medskip$%
\[%
{\textstyle\sum\nolimits_{0}}
=\left[
\begin{array}
[c]{cc}%
\dfrac{\alpha_{0}^{2}}{270}+\dfrac{\alpha_{0}\beta_{0}}{540}+\dfrac{\beta
_{0}^{2}}{3780}+\dfrac{1}{5} & \dfrac{\beta_{0}^{2}}{8640}+\dfrac{\alpha
_{0}\beta_{0}}{2160}\medskip\medskip\\
\dfrac{\beta_{0}^{2}}{8640}+\dfrac{\alpha_{0}\beta_{0}}{2160} &
\ \ \ \ \ \dfrac{\alpha_{0}^{2}}{105}+\dfrac{\alpha_{0}\beta_{0}}{252}%
+\dfrac{17\beta_{0}^{2}}{21000}+\dfrac{1}{15}%
\end{array}
\right]  ,
\]

\noindent it follows that$\medskip$%

\begin{align*}%
{\textstyle\sum\nolimits^{2}}
&  =\left[
\begin{array}
[c]{cc}%
\dfrac{342\alpha_{0}^{2}}{35}+\dfrac{327\alpha_{0}\beta_{0}}{70}%
+\dfrac{263\beta_{0}^{2}}{280}+\dfrac{624}{5} & \ \ \ \ \dfrac{240\alpha
_{0}^{2}}{7}+\dfrac{107\alpha_{0}\beta_{0}}{7}+\dfrac{443\beta_{0}^{2}}%
{140}+240\medskip\medskip\\
\dfrac{240\alpha_{0}^{2}}{7}+\dfrac{107\alpha_{0}\beta_{0}}{7}+\dfrac
{443\beta_{0}^{2}}{140}+240 & \ \ \ \ \dfrac{960\alpha_{0}^{2}}{7}%
+\dfrac{400\alpha_{0}\beta_{0}}{7}+\dfrac{408\beta_{0}^{2}}{35}+960
\end{array}
\right]  .\\
&
\end{align*}
Finally, we note that assumptions $\left[  H.1\right]  -\left[  H.3\right]  $
may be also verified for one and two parameters copula families given in
(\ref{gumbel1}) and (\ref{gumbel}), respectively, but that requires tedious
calculations which would get us out of the context of the paper.

\section{\label{simulation}\textbf{Simulation study}}

\noindent To evaluate and compare the performance of BLM's estimator with
other estimators a simulation study is carried out with $r=2$ by considering
$C_{\alpha_{1},\alpha_{2}}$ (the one iterated FGM family) and $C_{\beta
_{1},\beta_{2}}$ (the two parameters Gumbel family) given in $\left(
\ref{eq8}\right)  $ and $\left(  \ref{gumbel}\right)  $ respectively.\ The
evaluation of the performance is based on the bias and the RMSE defined as
follows:{\small {%
\begin{equation}
\text{Bias}=\frac{1}{N}\sum_{i=1}^{N}\left(  \hat{\theta}_{i}-\theta\right)
,\ \text{RMSE}=\left(  \frac{1}{N}\sum_{i=1}^{N}\left(  \hat{\theta}%
_{i}-\theta\right)  ^{2}\right)  ^{1/2}, \label{b}%
\end{equation}
}}where $\hat{\theta}_{i}$ is an estimator (from the considered method) of
$\theta$ from the $i$th samples for $N$ generated samples from the underlying
copula. In both parts, we selected $N=1000.$ We compare the BLM estimator with
the PML, $\left(  \tau,\rho\right)  -$inversion $\left(  \ref{rho-tau-sys}%
\right)  $ and MD estimators (see the Appendix). The procedure outlined in
Section (\ref{section4}) is repeated for different sample sizes $n$ with
$n=30,50,100,500$ to assess the improvement in the bias and RMSE of the
estimators with increasing sample size. Furthermore, the simulation procedure
is repeated for a large set of parameters of the true copulas $C_{\alpha
_{1},\alpha_{2}}$ and $C_{\beta_{1},\beta_{2}}.$ For each sample, we solve
systems $\left(  \ref{eq8}\right)  $ and $\left(  \ref{gumbelsystem}\right)  $
to obtain, respectively, the BLM-estimators $\left(  \widehat{\alpha}%
_{1,i},\widehat{\alpha}_{2,i}\right)  $ and $\left(  \widehat{\beta}%
_{1,i},\widehat{\beta}_{2,i}\right)  $ of $\left(  \alpha_{1},\alpha
_{2}\right)  $ and $\left(  \beta_{1},\beta_{2}\right)  $ for $i=1,...,N,$ and
the estimators $\widehat{\alpha}_{k},$ $\widehat{\beta}_{k}$ for $k=1,2$ are
given by $\widehat{\alpha}_{k}=\frac{1}{N}\sum_{i=1}^{N}\widehat{\alpha}%
_{k,i}$ and $\widehat{\beta}_{k}=\frac{1}{N}\sum_{i=1}^{N}\widehat{\beta
}_{k,i}.$

\subsection{Performance of the BLM-based estimation}

\noindent We first select parameters, as the true values of the parameters, of
Gumbel and FGM copula models. The choice of the parameters have to be
meaningful, in the sense that each couple of parameters assigns a value of one
of the dependence measure, that is weak, moderate and strong dependence.\ In
other words, if we consider Spearman's rho $\rho$ as a dependence measure,
then we should select values for copula parameters that correspond to
specified\ values of $\rho$ by using equation (\ref{rho}).\ Recall that for
the FGM family $C_{\alpha_{1},\alpha_{2}},$ the dependence reaches the maximum
$\rho_{FGM}^{\max}=0.42721$ in $\alpha_{1}=-1+7/\sqrt{13}\approx0.941$ and
$\alpha_{2}=2-2/\sqrt{13}\approx1.445$ (see (\ref{eq7*})). So, we may chose
$\left(  \alpha_{1},\alpha_{2}\right)  =\left(  0.941,1.445\right)  $ as the
true parameters of FGM family that correspond to the strong dependence. For
the true values of $\left(  \alpha_{1},\alpha_{2}\right)  $ corresponding to
the weak and the moderate dependence, we proceed as follows. We assign a value
to the couple $\left(  \rho,\alpha_{1}\right)  $ such that $\left\vert
\alpha_{1}\right\vert \leq1,$ then we solve by numerical methods the equation
(\ref{rho}) in the region (\ref{eq7}) and get the corresponding value to
$\alpha_{2}.$ We summarize the results in the following table:%

\begin{table}[h] \centering
$%
\begin{tabular}
[c]{c|c|c}%
$\rho$ & $\alpha_{1}$ & $\alpha_{2}$\\\hline\hline
$0.001$ & $0.100$ & $0$\\\hline
$0.208$ & $0.400$ & $0.900$\\\hline
$0.427$ & $0.941$ & $1.445$\\\hline
\end{tabular}
\ $%
\caption{The true parameters of FGM copula used for the simulation study .}\label{TAB1}%
\end{table}%

\noindent By the same procedure, we select the true parameters $\left(
\beta_{1},\beta_{2}\right)  $ of the Gumbel copula $C_{\beta_{1},\beta_{2}}$
and get:%

\begin{table}[h] \centering
$%
\begin{tabular}
[c]{c|c|c}%
$\rho$ & $\beta_{1}$ & $\beta_{2}$\\\hline\hline
$0.001$ & $1$ & $0.001$\\\hline
$0.500$ & $1.400$ & $0.200$\\\hline
$0.900$ & $2.500$ & $1$\\\hline
\end{tabular}
\ $%
\caption{The true parameters of Gumbel copula used for the simulation study.}\label{TAB2}%
\end{table}%

\noindent To evaluate the performance of the BLM estimators, we proceed as follows:

\begin{enumerate}
\item By using the Algorithm in Nelsen 2006 page 41 and Theorem 4.3.7 in
Nelsen 2006 page 129, respectively, we generate twice $N$ samples of size $n$
from each one the considered copulas $C_{\alpha_{1},\alpha_{2}}$ and
$C_{\beta_{1},\beta_{2}}.$

\item Obtain the BLM estimators $\left(  \widehat{\alpha}_{1},\widehat{\alpha
}_{2}\right)  $ of $\left(  \alpha_{1},\alpha_{2}\right)  $ and $\left(
\widehat{\beta}_{1},\widehat{\beta}_{2}\right)  $ of $\left(  \beta_{1}%
,\beta_{2}\right)  .$

\item By computing, for each estimator, the appropriate Bias and RMSE, we
compare $\left(  \widehat{\alpha}_{1},\widehat{\alpha}_{2}\right)  $ and
$\left(  \widehat{\beta}_{1},\widehat{\beta}_{2}\right)  ,$ respectively, with
the true parameters $\left(  \alpha_{1},\alpha_{2}\right)  $ and $\left(
\beta_{1},\beta_{2}\right)  .$
\end{enumerate}

\noindent All computations were performed in the R Software version
2.10.1.\ The results of the simulation study are summarized in Tables \ref{A1}
and \ref{A2}. \ We observed that BLM's method provides, in terms of bias and
RMSE, reasonable results, notably when the sample size increases.\ However, in
the case of strong dependence for FGM's family when the sample size is small
and less than $30$, the estimation of the first parameter $\alpha_{1}$ is
better that of the second one $\alpha_{2}.$ However, for the sample sizes
greater than $100$ the results become reasonable and more better for sample
sizes greater than $n=500.$ For Gumbel family the performance of BLM's method
looks good even for small samples.

\subsection{Comparative study: BLM, $\left(  \tau,\rho\right)  -$inversion, MD
and PML}

\noindent As the previous Subsection, we consider the bivariate two-parameter
FGM and Gumbel copula families with the trues parameters those given in Tables
\ref{TAB1} and \ref{TAB2} respectively. The simulation\ study proceeds as
follows:\medskip

\begin{enumerate}
\item Generate $N$ samples of size $n=30,$ $50,$ $100,$ $500$ from the copula
$C_{\theta}.$

\item Assess the performances of the BLM, $\left(  \tau,\rho\right)
-$inversion, MD and PML estimators.

\item Compare the BLM, $\left(  \tau,\rho\right)  -$inversion, MD and PML
estimators with the true parameter $\mathbf{\theta}$ by computing, for each
estimator, the appropriate criteria given by $\left(  \ref{b}\right)  .$
\end{enumerate}

\noindent It is clear, from Tables \ref{A3} to \ref{A8}, that the BLM estimate
preforms better than the $\left(  \tau,\rho\right)  -$inversion, MD and PML
ones as far as the Bias is concerned. On the other hand, in the case of small
samples the $\left(  \tau,\rho\right)  -$inversion, MD and PLM methods give
better RMSE than the BLM one. However, when the sample size increases, the
RMSE of the BLM estimator becomes reasonable. Moreover, for the computation
time point of view, we observed that the $\left(  \tau,\rho\right)
-$inversion, MD and PLM estimates require hours to be obtained, notably when
the sample size becomes large, whereas the BLM estimate execution time is in
terms of minutes. This is a natural conclusion, because the $\left(  \tau
,\rho\right)  -$inversion, MD and PLM methods use the optimization problem
under constraints, while the BLM method uses systems of equations.

\subsection{Comparative robustness study: BLM, $\left(  \tau,\rho\right)
-$inversion, MD and PML}

\noindent In this subsection we study the sensitivity to outliers of BLM's
estimator\ and compare with those of the $\left(  \tau,\rho\right)
-$inversion, MD and PML ones. We consider an $\epsilon-$contaminated model for
two-parameters FGM familly by means of a copula from the same family. In other
terms, we are dealing with the following mixture copula model:%
\begin{equation}
C_{\alpha_{1},\alpha_{2}}\left(  \epsilon\right)  :=\left(  1-\epsilon\right)
C_{\alpha_{1},\alpha_{2}}+\epsilon C_{\alpha_{1}^{\ast},\alpha_{2}^{\ast}%
},\text{ }\label{mixture}%
\end{equation}
where $0<\epsilon<1$ is the amount of contamination. \ For the implementation
of mixtures models to the study outliers one refers, for instance, to Barnett
and Lewis (1994), page 43. In this context, we proceed our study as follows.
First, we select $\left(  \alpha_{1},\alpha_{2}\right)  =\left(
0.4,0.9\right)  $ corresponds to Spearman's Rho $\rho=0.208$ (see Table
\ref{TAB1}) and chose $\left(  \alpha_{1}^{\ast},\alpha_{2}^{\ast}\right)
=\left(  0,0\right)  $ to have the contamination model as the product copula
that is $C_{\alpha_{1}^{\ast},\alpha_{2}^{\ast}}(u,v)=uv.$\ Then we consider
four contamination scenarios according to $\epsilon=5\%,10\%,20\%,30\%.\ $For
each value $\epsilon,$ we generate $1000$ samples of size $n=40$ from the
copula $C_{\alpha_{1},\alpha_{2}}\left(  \epsilon\right)  .$\ Finally, we
compare the BLM, $\left(  \tau,\rho\right)  -$inversion, MD and PML estimators
with the true parameter $\left(  \alpha_{1},\alpha_{2}\right)  $ by computing,
for each estimator, the appropriate Bias and RMSE and summarize the results in
Table \ref{A9}. We observed that, for example, in $0\%$ contamination the
$\left(  \text{Bias, RMSE}\right)  $ of $\widehat{\alpha}_{1}$ equals $\left(
0.044,0.832\right)  ,$ while for $30\%$ contamination is $\left(
-0.165,0.835\right)  .$ We may conclude that the RMSE of BLM's estimation in
less sensitive (or robust) to outliers, however the Bias is not. The same
conclusion is for the $\left(  \tau,\rho\right)  -$inversion method but the
BLM one is better.\ For PLM's estimation both the Bias and the RMSE are
sensitive, indeed for $0\%$ contamination the $\left(  \text{Bias,
RMSE}\right)  $ of $\widehat{\alpha}_{1}$ equals $\left(  -0.238,0.440\right)
,$ while for $30\%$ contamination is $\left(  -0.328,0.589\right)  .$ Both the
bias and the RMSE of MD's estimation are note sensitive to outliers, then we
may conclude that is the better among the four estimation methods. However,
the computation time cost in MD's method is important which is considered as
an handicap from practitioners.

\section{\textbf{Conclusions}}

\noindent In this paper, a formula of the bivariate \textit{L}-moments in
terms of copulas is given. This formula leads to introduce a new estimation
method for bivariate copula parameters, that we called the BLM based
estimation. The limiting distribution of the estimators given by the BLM
method are established.\ Moreover, we compared by simulations the BLM method
with the $\left(  \tau,\rho\right)  $-inversion, the minimum distance (MD) and
the pseudo maximum likelihood (PML) estimators by focusing on the Bias and the
RMSE. We conclude that the BLM based estimation performs well the Bias and
reasonably the RMSE. However, BLM's method may be an alternative robust method
as far as the RMSE is concerned. As final conclusion, it is worth noting that
computation's time of the proposed method is quite small compared to MD and
PML ones.\medskip

\noindent\textbf{Acknowledgements}\newline\noindent The authors are grateful
to Professor Taha B.M.J. Ouarda from INRS-ETE for his valuable comments and
suggestions that helped to improve the quality of this paper.

{\small \appendix}

\section{\textbf{Appendix}}

\subsection{Proof of Theorem \ref{TH1}}

\noindent The proof is straightforward and based on elementary calculation.
Indeed, since $\left(  F_{2}(X^{(2)})\right)  $ is $\left(  0,1\right)
-$uniform r.v., then copula's representation of the joint df of the pair of
r.v.'s $\left(  X^{(1)},\left(  F_{2}(X^{(2)})\right)  ^{\ell}\right)  $ is%
\[
D(u,v):=C\left(  F_{1}\left(  u\right)  ,v^{1/\ell}\right)  ,\text{ }%
\ell=1,2,...,k,
\]
then, the covariance of $\left(  X^{(1)},\left(  F_{2}(X^{(2)})\right)
^{\ell}\right)  $ equals%

\begin{align*}
Cov\left(  X^{(1)},\left(  F_{2}\left(  X^{(2)}\right)  \right)  ^{\ell
}\right)   &  =\int_{\mathbb{R}}\int_{\mathbb{I}}\left(  D(u,v)-v^{1/\ell
}F_{1}(u)\right)  dvdu\\
&  =\int_{\mathbb{R}}\int_{\mathbb{I}}\left(  C\left(  F_{1}(u),v^{1/\ell
}\right)  -v^{1/\ell}F_{1}(u)\right)  dvdu\\
&  =\int_{\mathbb{I}^{2}}\left(  C\left(  u,v\right)  -uv\right)  dv^{\ell
}dF_{1}^{-1}(u),
\end{align*}
it follows that
\[
Cov\left(  X^{(1)},P_{k}\left(  F_{2}\left(  X^{(2)}\right)  \right)  \right)
=\sum_{\ell=1}^{k}p_{\ell,k}Cov\left(  X^{(1)},\left(  F_{2}\left(
X^{(2)}\right)  \right)  ^{\ell}\right)  .
\]
Since%

\begin{align*}
\delta_{k\left[  12\right]  }  &  =Cov\left(  X^{(1)},P_{k}\left(
F_{2}\left(  X^{(2)}\right)  \right)  \right)  =\sum_{\ell=1}^{k}p_{\ell
,k}Cov\left(  F_{1}\left(  X^{(1)}\right)  ,\left(  F_{2}\left(
X^{(2)}\right)  \right)  ^{\ell}\right) \\
&  =\sum_{\ell=1}^{k}p_{\ell,k}\int_{\mathbb{I}^{2}}\left(  C\left(
u,v\right)  -uv\right)  dv^{\ell}du\\
&  =\int_{\mathbb{I}^{2}}\left(  C\left(  u,v\right)  -uv\right)
dP_{k}\left(  v\right)  du,
\end{align*}
as sought.\hfill$\Box$

\subsection{Proof of Theorem \ref{TH2}}

\noindent The existence of a sequence of consistent roots $\widehat
{\mathbf{\theta}}^{BLM}$ to (\ref{system1}) or (\ref{eq11}), may be checked by
using a similar argument as the proof of \ Theorem 1 in Tsukahara (2005).
Indeed, we have only to check the conditions in Theorem A.10.2 in Bickel
\textit{et al.} (1993). Since we are dealing with an asymptotic result, we may
consider that, for all large $n,$ without loss of generality, that the
empirical df $F_{j:n}$ and their rescaled version $F_{j:n}^{+}$ have a same
effect.\ Therefore throughout the proof, we will make use of $F_{j:n}$ instead
of $F_{j:n}^{+}.$ For convenience we set
\[
\Phi_{n}\left(  \mathbf{\theta}\right)  :=\frac{1}{n}\sum_{i=1}^{n}%
\mathbf{L}\left(  F_{1:n}\left(  X_{i}^{\left(  1\right)  }\right)
,F_{2:n}\left(  X_{i}^{\left(  2\right)  }\right)  ;\mathbf{\theta}\right)
\text{ and }\Phi\left(  \mathbf{\theta}\right)  :=\int_{\mathbb{I}^{2}%
}\mathbf{L}\left(  u_{1},u_{2};\mathbf{\theta}\right)  dC_{\mathbf{\theta}%
_{0}}\left(  u_{1},u_{2}\right)  .
\]
By assumption $\left[  H.2\right]  ,$ it is clear that the following
derivatives exist:%

\begin{align*}
\overset{\bullet}{\Phi}_{n}\left(  \mathbf{\theta}\right)   &  =\frac
{\partial\Phi_{n}\left(  \mathbf{\theta}\right)  }{\partial\mathbf{\theta}%
}=\frac{1}{n}\sum_{i=1}^{n}\overset{\bullet}{\mathbf{L}}\left(  F_{1:n}\left(
X_{i}^{\left(  1\right)  }\right)  ,F_{2:n}\left(  X_{i}^{\left(  2\right)
}\right)  ;\mathbf{\theta}\right)  ,\\
\overset{\bullet}{\Phi}\left(  \mathbf{\theta}\right)   &  =\frac{\partial
\Phi\left(  \mathbf{\theta}\right)  }{\partial\mathbf{\theta}}=\int
_{\mathbb{I}^{2}}\overset{\bullet}{\mathbf{L}}\left(  u_{1},u_{2}%
;\mathbf{\theta}\right)  dC_{\mathbf{\theta}_{0}}\left(  u_{1},u_{2}\right)  .
\end{align*}
Next, we verify that%
\begin{equation}
\sup\left\{  \left\vert \overset{\bullet}{\Phi}_{n}\left(  \mathbf{\theta
}\right)  -\overset{\bullet}{\Phi}\left(  \mathbf{\theta}\right)  \right\vert
:\left\vert \mathbf{\theta}-\mathbf{\theta}_{0}\right\vert <\epsilon
_{n}\right\}  \overset{\mathbf{P}}{\rightarrow}0,\text{ as }n\rightarrow
\infty, \label{sup}%
\end{equation}
for any real sequence $\epsilon_{n}\rightarrow0.$ By using the triangular
inequality we get
\[
\left\vert \overset{\bullet}{\Phi}_{n}\left(  \mathbf{\theta}\right)
-\overset{\bullet}{\Phi_{n}}\left(  \mathbf{\theta}_{0}\right)  \right\vert
\leq\frac{1}{n}\sum_{i=1}^{n}\left\vert \overset{\bullet}{\mathbf{L}}\left(
F_{1:n}\left(  X_{i}^{\left(  1\right)  }\right)  ,F_{2:n}\left(
X_{i}^{\left(  2\right)  }\right)  ;\mathbf{\theta}\right)  -\overset{\bullet
}{\mathbf{L}}\left(  F_{1:n}\left(  X_{i}^{\left(  1\right)  }\right)
,F_{2:n}\left(  X_{i}^{\left(  2\right)  }\right)  ;\mathbf{\theta}%
_{0}\right)  \right\vert .
\]
Since $\overset{\bullet}{\mathbf{L}}$ is continuous in $\mathbf{\theta,}$ then%
\[
\sup\left\{  \left\vert \overset{\bullet}{\mathbf{L}}\left(  F_{1:n}\left(
X_{i}^{\left(  1\right)  }\right)  ,F_{2:n}\left(  X_{i}^{\left(  2\right)
}\right)  ;\mathbf{\theta}\right)  -\overset{\bullet}{\mathbf{L}}\left(
F_{1:n}\left(  X_{i}^{\left(  1\right)  }\right)  ,F_{2:n}\left(
X_{i}^{\left(  2\right)  }\right)  ;\mathbf{\theta}_{0}\right)  \right\vert
:\left\vert \mathbf{\theta}-\mathbf{\theta}_{0}\right\vert <\epsilon
_{n}\right\}  =o_{\mathbf{P}}\left(  1\right)  ,
\]
therefore
\begin{equation}
\sup\left\{  \left\vert \overset{\bullet}{\Phi}_{n}\left(  \mathbf{\theta
}\right)  -\overset{\bullet}{\Phi_{n}}\left(  \mathbf{\theta}_{0}\right)
\right\vert :\left\vert \mathbf{\theta}-\mathbf{\theta}_{0}\right\vert
<\epsilon_{n}\right\}  \overset{\mathbf{P}}{\rightarrow}0,\text{ as
}n\rightarrow\infty. \label{sup2}%
\end{equation}
On the other hand, from the law of the large number, we infer that%
\[
\frac{1}{n}\sum_{i=1}^{n}\overset{\bullet}{\mathbf{L}}\left(  F_{1}\left(
X_{i}^{\left(  1\right)  }\right)  ,F_{2}\left(  X_{i}^{\left(  2\right)
}\right)  ;\mathbf{\theta}_{0}\right)  \overset{\mathbf{P}}{\rightarrow
}\overset{\bullet}{\Phi}\left(  \mathbf{\theta}_{0}\right)  ,\text{ as
}n\rightarrow\infty.
\]
Moreover, in view of the continuity of $\overset{\bullet}{\mathbf{L}}$ in $u$
and since $\sup_{x^{\left(  j\right)  }}\left\vert F_{i:n}\left(  x^{\left(
j\right)  }\right)  -F_{i}\left(  x^{\left(  j\right)  }\right)  \right\vert
\rightarrow0,$ $j=1,2,$ almost surely, $n\rightarrow\infty$ (Glivenko-Cantelli
theorem), we have%

\[
\frac{1}{n}\sum_{i=1}^{n}\left\vert \overset{\bullet}{\mathbf{L}}\left(
F_{1:n}\left(  X_{i}^{\left(  1\right)  }\right)  ,F_{2:n}\left(
X_{i}^{\left(  2\right)  }\right)  ;\mathbf{\theta}_{0}\right)  -\overset
{\bullet}{\mathbf{L}}\left(  F_{1}\left(  X_{i}^{\left(  1\right)  }\right)
,F_{2}\left(  X_{i}^{\left(  2\right)  }\right)  ;\mathbf{\theta}_{0}\right)
\right\vert \overset{\mathbf{P}}{\rightarrow}0.
\]

\noindent It follows that $\left\vert \overset{\bullet}{\Phi}_{n}\left(
\mathbf{\theta}_{0}\right)  -\overset{\bullet}{\Phi}\left(  \mathbf{\theta
}_{0}\right)  \right\vert \overset{\mathbf{P}}{\rightarrow}0,$ which together
with (\ref{sup2}), implies (\ref{sup}).$\medskip$

\noindent Conditions (MG0) and (MG3) in Theorem A.10.2 in Bickel \textit{et
al.} (1993) are trivially satisfied by our assumptions $\left[  H1\right]
-\left[  H3\right]  .$ In view of the general theorem for $Z$-estimators (see,
van der Vaart and Wellner, 1996, Th. 3.3.1), it remains to prove that
$\sqrt{n}\left(  \overset{\bullet}{\Phi}_{n}-\overset{\bullet}{\Phi}\right)
\left(  \mathbf{\theta}_{0}\right)  $ converges in law to the appropriate
limit. But this follows from Proposition 3 in Tsukahara (2005), which achieves
the proof of Theorem \ref{TH2}.\hfill$\Box$

\subsection{Minimum distance based estimation}

\noindent We briefly present the minimum distance (MD) base estimation for
copula models that possesses a qualitative robustness (Genest and
R\'{e}millard, 2008), this will be compared with the BLM method (see
Subsection 5.2). $\medskip$ \noindent Let $C$ be the true copula associated to
the df of $\left(  X^{\left(  1\right)  },X^{\left(  2\right)  }\right)  $ and
suppose that we have a given parametric family of copula $\mathcal{C}%
:=\left\{  C_{\mathbf{\theta}},\text{ }\mathbf{\theta}\in\mathcal{O}\right\}
$ to fit data.\ Let us define the minimum distance functional $T$ on the space
of the copula by%
\[
T\left(  C\right)  :=\arg\min_{\mathbf{\theta\in}\mathcal{O}}\mu\left(
C,C_{\mathbf{\theta}}\right)  .
\]
Here $\mu$ is a distance between probabilities on $\mathbb{I}^{2}.$ In the
present paper, we consider the Cram\'{e}r-von Mises distance defined by%
\[
\mu^{CVM}\left(  C,C_{\mathbf{\theta}}\right)  :=\int_{\mathbb{I}^{2}}\left\{
C\left(  u_{1},u_{2}\right)  -C_{\mathbf{\theta}}\left(  u_{1},u_{2}\right)
\right\}  ^{2}dC\left(  u_{1},u_{2}\right)  .
\]
\noindent Consider now a random sample $\left(  X_{i}^{(1)},X_{i}%
^{(2)}\right)  _{i=1,n},$ from the bivariate random variables $\left(
X^{(1)},X^{(2)}\right)  .$ The joint empirical distribution functions is given
by%
\[
F_{n}\left(  x_{1},x_{2}\right)  =\frac{1}{n}\sum_{i=1}^{n}\mathbf{1}\left\{
X_{i}^{(1)}\leq x_{1},X_{i}^{(2)}\leq x_{2}\right\}  .
\]
Following Deheuvels (1979), we define the empirical copula by%
\[
C_{n}\left(  u_{1},u_{2}\right)  :=F_{n}\left(  F_{n:1}^{-1}\left(
u_{1}\right)  ,F_{n:2}^{-1}\left(  u_{2}\right)  \right)  ,\text{ }0\leq
u_{1},u_{2}\leq1.
\]
The corresponding Cram\'{e}r-von Mises statistics is
\[
\mu^{CVM}\left(  C_{n},C_{\mathbf{\theta}}\right)  =\int_{\mathbb{I}^{2}%
}\left\{  C_{n}\left(  u_{1},u_{2}\right)  -C_{\mathbf{\theta}}\left(
u_{1},u_{2}\right)  \right\}  ^{2}dC_{n}\left(  u_{1},u_{2}\right)  .
\]
This may be rewritten into%
\[
\mu^{CVM}\left(  C_{n},C_{\mathbf{\theta}}\right)  =n^{-1}\sum_{i=1}%
^{n}\left(  C_{n}\left(  \widehat{U}_{i}^{\left(  1\right)  },\widehat{U}%
_{i}^{\left(  2\right)  }\right)  -C_{\mathbf{\theta}}\left(  \widehat{U}%
_{i}^{\left(  1\right)  },\widehat{U}_{i}^{\left(  2\right)  }\right)
\right)  ^{2},
\]
where $\widehat{U}_{i}^{\left(  j\right)  }:=F_{j:n}^{\ast}\left(
X_{i}^{\left(  j\right)  }\right)  ,$ $i=1,...,n,$ for each $j=1,2$ (see,
Genest and R\'{e}millard, 2008, eq.\ 31).\ The MD estimator of the parameter
$\mathbf{\theta}$ is defined by%
\[
\widehat{\mathbf{\theta}}=T\left(  C_{n}\right)  :=\arg\min_{\mathbf{\theta
\in}\mathcal{O}}\mu^{CVM}\left(  C_{n},C_{\mathbf{\theta}}\right)  .
\]
Note that we may also use the Kolmogorov-Smirnov distance but this is awkward
in practice due to the supremum norm uses. Also since the Hellinger distance
is defined by copula densities, other nonparametric estimators of the
underling copula are needed (see, Biau and Begkamp, 2005) and therefore
non-standard computational procedures are required.\medskip

\noindent Suppose now that we are dealing with the estimation of parameters of
one iterated FGM copula family $C_{\alpha_{1},\alpha_{2}}$ in $\left(
\ref{eq6}\right)  $ by means of the MD method. The MD estimator for
$\mathbf{\alpha=}\left(  \alpha_{1},\alpha_{2}\right)  $ noted $\widehat
{\mathbf{\alpha}}^{MD}$ results by minimizing the function $\left(  \alpha
_{1},\alpha_{2}\right)  \rightarrow\rho\left(  C_{n},C_{\alpha_{1},\alpha_{2}%
}\right)  $ over the region $\mathcal{R}$ given in $\left(  \ref{eq7}\right)
.$ Then to solve the previous optimization problem, we will introduce the
Lagrange multiplier principle, that is we have to rewrite the region
$\mathcal{R}$ into
\[
\mathcal{R=}\left\{  \left(  \alpha_{1},\alpha_{2}\right)  ,\text{ }\ell
_{j}\left(  \alpha_{1},\alpha_{2}\right)  \geq0,\text{ }j=1,2,3\right\}  ,
\]
where $\ell_{1}\left(  \alpha_{1},\alpha_{2}\right)  :=1-\alpha_{1}^{2},$
$\ell_{2}\left(  \alpha_{1},\alpha_{2}\right)  :=\alpha_{1}+\alpha_{2}+1$ and
\[
\ell_{3}\left(  \alpha_{1},\alpha_{2}\right)  :=\frac{1}{2}\left[
3-\alpha_{1}+\left(  9-6\alpha_{1}-3\alpha_{1}^{2}\right)  ^{1/2}\right]
-\alpha_{2},
\]
and then minimize the function%
\[
\mathbb{K}_{n}\left(  \mathbf{\alpha,\nu}\right)  :=\rho\left(  C_{n}%
,C_{\alpha_{1},\alpha_{2}}\right)  -\sum_{j=1}^{3}\nu_{j}\ell_{j}\left(
\alpha_{1},\alpha_{2}\right)  ,
\]
over the whole $\mathbb{R}^{5},$ with $\mathbf{\alpha=}\left(  \alpha
_{1},\alpha_{2}\right)  \in\mathbb{R}^{2}$ and $\mathbf{\nu=}\left(  \nu
_{1},\nu_{2},\nu_{3}\right)  \in\mathbb{R}^{3}.$ \ So, the new formulation of
the MD estimator of parameter $\mathbf{\alpha}$ is%
\[
\widehat{\mathbf{\alpha}}^{MD}=\arg\min_{\left(  \mathbf{\alpha,\nu}\right)
\in\mathbb{R}^{5}}\mathbb{K}_{n}\left(  \mathbf{\alpha,\nu}\right)  .
\]
We note here that it is difficult, in general, to have an explicit form for
$\widehat{\mathbf{\alpha}}^{MD},$ then only the numerical computation can
solve this issue. This is observed for the one-iterated FGM family, that the
optimization problem requires tedious tools.%

\begin{landscape}%
\[
\]
%

\begin{table}[h] \centering
$%
\begin{tabular}
[c]{ccccccccccccc}
& \multicolumn{4}{c}{$\rho=0.001$} & \multicolumn{4}{|c}{$\rho=0.208$} &
\multicolumn{4}{|c}{$\rho=0.427$}\\\hline
& \multicolumn{2}{c}{$\alpha_{1}=0.1$} & \multicolumn{2}{c}{$\alpha_{2}=0$} &
\multicolumn{2}{c}{$\alpha_{1}=0.4$} & \multicolumn{2}{c}{$\alpha_{2}=0.9$} &
\multicolumn{2}{c}{$\alpha_{1}=0.941$} & \multicolumn{2}{c}{$\alpha_{2}%
=1.445$}\\\hline
\multicolumn{1}{l}{$n${\small \ \ }} & {\small Bias} & {\small RMSE} &
{\small Bias} & {\small RMSE} & \multicolumn{1}{|c}{{\small Bias}} &
{\small RMSE} & {\small Bias} & {\small RMSE} &
\multicolumn{1}{|c}{{\small Bias}} & {\small RMSE} & {\small Bias} &
{\small RMSE}\\\hline\hline
\multicolumn{1}{l}{${\small 30}$} & \multicolumn{1}{r}{${\small 0.197}$} &
\multicolumn{1}{r}{${\small 0.882}$} & \multicolumn{1}{r}{${\small -0.089}$} &
\multicolumn{1}{r|}{${\small 2.417}$} & \multicolumn{1}{|c}{${\small 0.113}$}
& \multicolumn{1}{r}{${\small 0.931}$} & ${\small -0.210}$ & ${\small 2.804}$
& \multicolumn{1}{|c}{${\small 0.098}$} & ${\small 0.823}$ & ${\small -0.312}$
& ${\small 2.861}$\\
\multicolumn{1}{l}{${\small 50}$} & \multicolumn{1}{r}{${\small 0.133}$} &
\multicolumn{1}{r}{${\small 0.672}$} & \multicolumn{1}{r}{${\small -0.050}$} &
\multicolumn{1}{r|}{${\small 1.781}$} & \multicolumn{1}{|r}{${\small 0.042}$}
& \multicolumn{1}{r}{${\small 0.703}$} & \multicolumn{1}{r}{${\small -0.065}$}
& \multicolumn{1}{r|}{${\small 2.074}$} & \multicolumn{1}{|c}{${\small -0.051}%
$} & ${\small 0.712}$ & ${\small 0.276}$ & ${\small 2.197}$\\
\multicolumn{1}{l}{${\small 100}$} & \multicolumn{1}{r}{${\small 0.065}$} &
\multicolumn{1}{r}{${\small 0.456}$} & \multicolumn{1}{r}{${\small -0.040}$} &
\multicolumn{1}{r|}{${\small 1.105}$} & \multicolumn{1}{|r}{${\small 0.026}$}
& \multicolumn{1}{r}{${\small 0.498}$} & \multicolumn{1}{r}{${\small 0.048}$}
& \multicolumn{1}{r|}{${\small 1.408}$} & \multicolumn{1}{|c}{${\small 0.041}%
$} & ${\small 0.513}$ & ${\small -0.055}$ & ${\small 1.572}$\\
\multicolumn{1}{l}{${\small 500}$} & \multicolumn{1}{r}{${\small -0.017}$} &
\multicolumn{1}{r}{${\small 0.206}$} & \multicolumn{1}{r}{${\small 0.041}$} &
\multicolumn{1}{r|}{${\small 0.639}$} & \multicolumn{1}{|r}{${\small 0.021}$}
& \multicolumn{1}{r}{${\small 0.215}$} & \multicolumn{1}{r}{${\small 0.031}$}
& \multicolumn{1}{r|}{${\small 0.659}$} & \multicolumn{1}{|c}{${\small -0.020}%
$} & ${\small 0.308}$ & ${\small -0.031}$ & ${\small 0.692}$\\\hline\hline
&  &  &  &  &  &  &  &  &  &  &  &
\end{tabular}
\ \ \ $\caption{Bias and RMSE of BLM's estimator of
two-parameters FGM copula.}\label{A1}%
\end{table}%
%

\begin{table}[h] \centering
$%
\begin{tabular}
[c]{ccccccccccccc}
& \multicolumn{4}{c}{$\rho=0.001$} & \multicolumn{4}{|c}{$\rho=0.5$} &
\multicolumn{4}{|c}{$\rho=0.9$}\\\hline
& \multicolumn{2}{c}{$\beta_{1}=1$} & \multicolumn{2}{c}{$\beta_{2}=0.001$} &
\multicolumn{2}{|c}{$\beta_{1}=1.4$} & \multicolumn{2}{c}{$\beta_{2}=0.2$} &
\multicolumn{2}{|c}{$\beta_{1}=2.5$} & \multicolumn{2}{c}{$\beta_{2}=1$%
}\\\hline
\multicolumn{1}{l}{$n${\small \ \ }} & {\small Bias} & {\small RMSE} &
{\small Bias} & {\small RMSE} & \multicolumn{1}{|c}{{\small Bias}} &
{\small RMSE} & {\small Bias} & {\small RMSE} &
\multicolumn{1}{|c}{{\small Bias}} & {\small RMSE} & {\small Bias} &
{\small RMSE}\\\hline\hline
\multicolumn{1}{l}{${\small 30}$} & \multicolumn{1}{r}{${\small 0.162}$} &
\multicolumn{1}{r}{${\small 0.994}$} & \multicolumn{1}{r}{${\small 0.428}$} &
\multicolumn{1}{r|}{${\small 1.945}$} & \multicolumn{1}{|r}{${\small 0.214}$}
& \multicolumn{1}{r}{${\small 1.002}$} & \multicolumn{1}{r}{${\small 0.549}$}
& \multicolumn{1}{r|}{${\small 1.421}$} & \multicolumn{1}{|r}{${\small 0.404}%
$} & \multicolumn{1}{r}{${\small 0.920}$} &
\multicolumn{1}{r}{${\small -0.653}$} & \multicolumn{1}{r}{${\small 1.109}$}\\
\multicolumn{1}{l}{${\small 50}$} & \multicolumn{1}{r}{${\small 0.134}$} &
\multicolumn{1}{r}{${\small 0.725}$} & \multicolumn{1}{r}{${\small 0.294}$} &
\multicolumn{1}{r|}{${\small 1.107}$} & \multicolumn{1}{|r}{${\small 0.187}$}
& \multicolumn{1}{r}{${\small 0.695}$} & \multicolumn{1}{r}{${\small 0.498}$}
& \multicolumn{1}{r|}{${\small 0.999}$} & \multicolumn{1}{|r}{${\small 0.350}%
$} & \multicolumn{1}{r}{${\small 0.854}$} &
\multicolumn{1}{r}{${\small -0.550}$} & \multicolumn{1}{r}{${\small 0.835}$}\\
\multicolumn{1}{l}{${\small 100}$} & \multicolumn{1}{r}{${\small -0.094}$} &
\multicolumn{1}{r}{${\small 0.697}$} & \multicolumn{1}{r}{${\small 0.219}$} &
\multicolumn{1}{r|}{${\small 0.804}$} & \multicolumn{1}{|r}{${\small -0.136}$}
& \multicolumn{1}{r}{${\small 0.619}$} & \multicolumn{1}{r}{${\small 0.287}$}
& \multicolumn{1}{r|}{${\small 0.665}$} & \multicolumn{1}{|r}{${\small 0.183}%
$} & \multicolumn{1}{r}{${\small 0.597}$} &
\multicolumn{1}{r}{${\small -0.536}$} & \multicolumn{1}{r}{${\small 0.526}$}\\
\multicolumn{1}{l}{${\small 500}$} & \multicolumn{1}{r}{${\small -0.071}$} &
\multicolumn{1}{r}{${\small 0.597}$} & \multicolumn{1}{r}{${\small 0.107}$} &
\multicolumn{1}{r|}{${\small 0.358}$} & \multicolumn{1}{|r}{${\small -0.081}$}
& \multicolumn{1}{r}{${\small 0.489}$} & \multicolumn{1}{r}{${\small 0.148}$}
& \multicolumn{1}{r|}{${\small 0.477}$} & \multicolumn{1}{|r}{${\small -0.096}%
$} & \multicolumn{1}{r}{${\small 0.395}$} &
\multicolumn{1}{r}{${\small -0.340}$} & \multicolumn{1}{r}{${\small 0.480}$%
}\\\hline\hline
&  &  &  &  &  &  &  &  &  &  &  &
\end{tabular}
\ $%
\caption{Bias and RMSE of BLM's estimator of two-parameters FGM copula.}\label{A2}%
\end{table}%
%

\end{landscape}%
{\tiny
\begin{table}[h] \centering
$%
\begin{tabular}
[c]{lrlrrl}
& \multicolumn{2}{c}{$\alpha_{1}=0.1$} & \multicolumn{2}{c}{$\alpha_{2}=0$} &
\\\hline
\ \ \ \ \ \ \ \ \ \ \ \ \ \ \ \  & \ \ \ \ \ \ \ Bias & \ \ \ \ \ RMSE &
\ \ \ \ \ \ Bias & \ \ RMSE & \multicolumn{1}{c}{{\tiny Time (h)}%
}\\\hline\hline
\multicolumn{6}{c}{$n=30$}\\\hline
BLM & $0.227$ & \multicolumn{1}{r}{$0.952$} & $-0.194$ & $2.882$ &
\multicolumn{1}{r}{$0.640$}\\
$\left(  \tau,\rho\right)  $-inversion & $0.458$ & \multicolumn{1}{r}{$1.005$}
& $0.782$ & $2.157$ & \multicolumn{1}{r}{$0.978$}\\
MD & $0.575$ & \multicolumn{1}{r}{$0.571$} & $0.494$ & $0.851$ &
\multicolumn{1}{r}{$1.566$}\\
PML & $0.550$ & \multicolumn{1}{r}{$0.552$} & $0.424$ & $0.872$ &
\multicolumn{1}{r}{$1.033$}\\\hline
\multicolumn{6}{c}{$n=50$}\\\hline
BLM & $-0.140$ & \multicolumn{1}{r}{$0.702$} & $-0.112$ & $2.193$ &
\multicolumn{1}{r}{$1.215$}\\
$\left(  \tau,\rho\right)  $-inversion & $0.358$ & \multicolumn{1}{r}{$0.958$}
& $0.558$ & $1.428$ & \multicolumn{1}{r}{$2.856$}\\
MD & $0.468$ & \multicolumn{1}{r}{$0.559$} & $0.238$ & $0.846$ &
\multicolumn{1}{r}{$3.455$}\\
PML & $0.444$ & \multicolumn{1}{r}{$0.546$} & $-0.237$ & $0.840$ &
\multicolumn{1}{r}{$2.421$}\\\hline
\multicolumn{6}{c}{$n=100$}\\\hline
BLM & $-0.039$ & \multicolumn{1}{r}{$0.565$} & $0.082$ & $1.364$ &
\multicolumn{1}{r}{$1.847$}\\
$\left(  \tau,\rho\right)  $-inversion & $0.229$ & \multicolumn{1}{r}{$0.664$}
& $0.195$ & $0.985$ & \multicolumn{1}{r}{$2.548$}\\
MD & $0.125$ & \multicolumn{1}{r}{$0.521$} & $0.145$ & $0.684$ &
\multicolumn{1}{r}{$6.888$}\\
PML & $0.121$ & \multicolumn{1}{r}{$0.520$} & $0.131$ & $0.673$ &
\multicolumn{1}{r}{$4.107$}\\\hline
\multicolumn{6}{c}{$n=500$}\\\hline
BLM & $0.021$ & \multicolumn{1}{r}{$0.417$} & $0.071$ & $0.634$ &
\multicolumn{1}{r}{$5.963$}\\
$\left(  \tau,\rho\right)  $-inversion & $0.084$ & \multicolumn{1}{r}{$0.588$}
& $0.118$ & $0.748$ & \multicolumn{1}{r}{$11.548$}\\
MD & $0.077$ & \multicolumn{1}{r}{$0.504$} & $0.086$ & $0.640$ &
\multicolumn{1}{r}{$19.598$}\\
PML & $0.076$ & \multicolumn{1}{r}{$0.502$} & $0.081$ & $0.639$ &
\multicolumn{1}{r}{$17.073$}\\\hline\hline
&  &  &  &  &
\end{tabular}
\ \ \ \ \ \ \ \ \ \ $%
\caption{Bias and RMSE of the BLM, ($\tau$,$\rho$)-inversion, MD and PML
estimators for two-parameters of FGM copula for weak dependence ($\rho=0.001$).}\label{A3}%
\end{table}%
}

{\tiny
\begin{table}[h] \centering
$%
\begin{tabular}
[c]{lrlrrl}
& \multicolumn{2}{c}{$\alpha_{1}=0.4$} & \multicolumn{2}{c}{$\alpha_{2}=0.9$}
& \\\hline
\ \ \ \ \ \ \ \ \ \ \ \ \ \ \ \  & \ \ \ \ \ \ \ Bias & \ \ \ \ \ RMSE &
\ \ \ \ \ \ Bias & \ \ RMSE & \multicolumn{1}{c}{{\tiny Time (h)}%
}\\\hline\hline
\multicolumn{6}{c}{$n=30$}\\\hline
BLM & $0.127$ & \multicolumn{1}{r}{$0.855$} & $-0.297$ & $2.668$ &
\multicolumn{1}{r}{$0.011$}\\
$\left(  \tau,\rho\right)  $-inversion & $0.102$ & \multicolumn{1}{r}{$1.322$}
& $-0.290$ & $1.383$ & \multicolumn{1}{r}{$0.516$}\\
MD & $-0.174$ & \multicolumn{1}{r}{$0.777$} & $-0.322$ & $1.058$ &
\multicolumn{1}{r}{$3.583$}\\
PML & $-0.191$ & \multicolumn{1}{r}{$0.906$} & $-0.372$ & $1.261$ &
\multicolumn{1}{r}{$0.954$}\\\hline
\multicolumn{6}{c}{$n=50$}\\\hline
BLM & $-0.059$ & \multicolumn{1}{r}{$0.755$} & $0.123$ & $2.001$ &
\multicolumn{1}{r}{$1.035$}\\
$\left(  \tau,\rho\right)  $-inversion & $0.091$ & \multicolumn{1}{r}{$0.892$}
& $-0.141$ & $1.272$ & \multicolumn{1}{r}{$2.101$}\\
MD & $-0.173$ & \multicolumn{1}{r}{$0.730$} & $-0.223$ & $1.010$ &
\multicolumn{1}{r}{$6.428$}\\
PML & $0.122$ & \multicolumn{1}{r}{$0.775$} & $-0.200$ & $0.853$ &
\multicolumn{1}{r}{$2.823$}\\\hline
\multicolumn{6}{c}{$n=100$}\\\hline
BLM & $0.031$ & \multicolumn{1}{r}{$0.715$} & $0.060$ & $1.404$ &
\multicolumn{1}{r}{$1.920$}\\
$\left(  \tau,\rho\right)  $-inversion & $0.082$ & \multicolumn{1}{r}{$0.791$}
& $-0.130$ & $0.942$ & \multicolumn{1}{r}{$2.037$}\\
MD & $-0.130$ & \multicolumn{1}{r}{$0.652$} & $-0.121$ & $0.919$ &
\multicolumn{1}{r}{$11.217$}\\
PML & $0.090$ & \multicolumn{1}{r}{$0.599$} & $0.100$ & $0.794$ &
\multicolumn{1}{r}{$3.652$}\\\hline
\multicolumn{6}{c}{$n=500$}\\\hline
BLM & $-0.025$ & \multicolumn{1}{r}{$0.300$} & $0.049$ & $0.629$ &
\multicolumn{1}{r}{$5.205$}\\
$\left(  \tau,\rho\right)  $-inversion & $0.054$ & \multicolumn{1}{r}{$0.393$}
& $-0.087$ & $0.701$ & \multicolumn{1}{r}{$8.285$}\\
MD & $-0.071$ & \multicolumn{1}{r}{$0.602$} & $-0.061$ & $0.742$ &
\multicolumn{1}{r}{$19.210$}\\
PML & $0.047$ & \multicolumn{1}{r}{$0.573$} & $0.056$ & $0.632$ &
\multicolumn{1}{r}{$16.458$}\\\hline\hline
&  &  &  &  &
\end{tabular}
\ \ \ \ \ \ \ \ \ \ \ \ \ \ $%
\caption{Bias and RMSE of the BLM, ($\tau$,$\rho$)-inversion, MD and PML
estimators for two-parameters of FGM copula for moderate dependence ($\rho=0.208$).}\label{A4}%
\end{table}%
}

{\tiny
\begin{table}[h] \centering
$%
\begin{tabular}
[c]{lrlrrl}
& \multicolumn{2}{c}{$\alpha_{1}=0.941$} & \multicolumn{2}{c}{$\alpha
_{2}=1.445$} & \\\hline
\ \ \ \ \ \ \ \ \ \ \ \ \ \ \ \  & \ \ \ \ \ \ \ Bias & \ \ \ \ \ RMSE &
\ \ \ \ \ \ Bias & \ \ RMSE & {\tiny Time (h)}\\\hline\hline
\multicolumn{6}{c}{$n=30$}\\\hline
BLM & $0.091$ & \multicolumn{1}{r}{$0.832$} & $0.402$ & $2.715$ &
\multicolumn{1}{r}{$0.017$}\\
$\left(  \tau,\rho\right)  $-inversion & $0.171$ & \multicolumn{1}{r}{$1.142$}
& $-0.471$ & $1.229$ & \multicolumn{1}{r}{$0.757$}\\
MD & $-0.142$ & \multicolumn{1}{r}{$0.871$} & $-0.420$ & $1.025$ &
\multicolumn{1}{r}{$2.083$}\\
PML & $-0.121$ & \multicolumn{1}{r}{$0.927$} & $-0.415$ & $1.061$ &
\multicolumn{1}{r}{$1.781$}\\\hline
\multicolumn{6}{c}{$n=50$}\\\hline
BLM & $0.054$ & \multicolumn{1}{r}{$0.641$} & $0.300$ & $1.982$ &
\multicolumn{1}{r}{$1.020$}\\
$\left(  \tau,\rho\right)  $-inversion & $0.157$ & \multicolumn{1}{r}{$0.997$}
& $-0.321$ & $1.120$ & \multicolumn{1}{r}{$2.021$}\\
MD & $-0.135$ & \multicolumn{1}{r}{$0.753$} & $0.351$ & $0.940$ &
\multicolumn{1}{r}{$6.633$}\\
PML & $0.092$ & \multicolumn{1}{r}{$0.892$} & $-0.307$ & $1.150$ &
\multicolumn{1}{r}{$2.754$}\\\hline
\multicolumn{6}{c}{$n=100$}\\\hline
BLM & $0.030$ & \multicolumn{1}{r}{$0.449$} & $0.090$ & $1.391$ &
\multicolumn{1}{r}{$1.620$}\\
$\left(  \tau,\rho\right)  $-inversion & $0.081$ & \multicolumn{1}{r}{$0.463$}
& $-0.153$ & $0.931$ & \multicolumn{1}{r}{$3.037$}\\
MD & $0.070$ & \multicolumn{1}{r}{$0.743$} & $-0.114$ & $0.904$ &
\multicolumn{1}{r}{$9.217$}\\
PML & $0.050$ & \multicolumn{1}{r}{$0.712$} & $0.102$ & $0.800$ &
\multicolumn{1}{r}{$3.652$}\\\hline
\multicolumn{6}{c}{$n=500$}\\\hline
BLM & $0.021$ & \multicolumn{1}{r}{$0.315$} & $0.046$ & $0.602$ &
\multicolumn{1}{r}{$5.205$}\\
$\left(  \tau,\rho\right)  $-inversion & $0.071$ & \multicolumn{1}{r}{$0.357$}
& $-0.098$ & $0.765$ & \multicolumn{1}{r}{$8.285$}\\
MD & $-0.064$ & \multicolumn{1}{r}{$0.541$} & $-0.054$ & $0.782$ &
\multicolumn{1}{r}{$19.210$}\\
PML & $0.052$ & \multicolumn{1}{r}{$0.472$} & $0.076$ & $0.699$ &
\multicolumn{1}{r}{$17.458$}\\\hline\hline
&  &  &  &  &
\end{tabular}
\ \ \ \ \ \ \ \ \ \ \ \ \ \ \ $%
\caption{Bias and RMSE of the BLM, ($\tau$,$\rho$)-inversion, MD and PML
estimators for two-parameters of FGM copula for strong dependence ($\rho=0.427$).}\label{A5}%
\end{table}%
}

{\tiny
\begin{table}[h] \centering
$%
\begin{tabular}
[c]{lrlrrl}
& \multicolumn{2}{c}{$\beta_{1}=1$} & \multicolumn{2}{c}{$\beta_{2}=0.001$} &
\\\hline
\ \ \ \ \ \ \ \ \ \ \ \ \ \ \ \  & \ \ \ \ \ \ \ Bias & \ \ \ \ \ RMSE &
\ \ \ \ \ \ Bias & \ \ RMSE & {\tiny Time (h)}\\\hline\hline
\multicolumn{6}{c}{$n=30$}\\\hline
BLM & $0.174$ & \multicolumn{1}{r}{$0.941$} & $0.453$ & $1.854$ &
\multicolumn{1}{c}{$1.121$}\\
$\left(  \tau,\rho\right)  $-inversion & $0.181$ & \multicolumn{1}{r}{$0.782$}
& $0.532$ & $1.186$ & \multicolumn{1}{c}{$2.021$}\\
MD & $-0.274$ & \multicolumn{1}{r}{$0.546$} & $-0.698$ & $1.243$ &
\multicolumn{1}{c}{$4.691$}\\
PML & $0.310$ & \multicolumn{1}{r}{$0.335$} & $-0.593$ & $0.910$ &
\multicolumn{1}{c}{$2.065$}\\\hline
\multicolumn{6}{c}{$n=50$}\\\hline
BLM & $-0.157$ & \multicolumn{1}{r}{$0.897$} & $0.289$ & $0.977$ &
\multicolumn{1}{c}{$1.026$}\\
$\left(  \tau,\rho\right)  $-inversion & $0.184$ & \multicolumn{1}{r}{$0.539$}
& $0.476$ & $0.629$ & \multicolumn{1}{c}{$2.265$}\\
MD & $0.262$ & \multicolumn{1}{r}{$0.448$} & $-0.310$ & $0.759$ &
\multicolumn{1}{c}{$5.633$}\\
PML & $0.250$ & \multicolumn{1}{r}{$0.303$} & $-0.302$ & $0.815$ &
\multicolumn{1}{c}{$2.754$}\\\hline
\multicolumn{6}{c}{$n=100$}\\\hline
BLM & $-0.126$ & \multicolumn{1}{r}{$0.530$} & $0.193$ & $0.824$ &
\multicolumn{1}{c}{$1.920$}\\
$\left(  \tau,\rho\right)  $-inversion & $-0.177$ & \multicolumn{1}{r}{$0.523$%
} & $0.250$ & $0.619$ & \multicolumn{1}{c}{$2.248$}\\
MD & $-0.161$ & \multicolumn{1}{r}{$0.420$} & $-0.201$ & $0.521$ &
\multicolumn{1}{c}{$6.285$}\\
PML & $0.151$ & \multicolumn{1}{r}{$0.272$} & $-0.197$ & $0.810$ &
\multicolumn{1}{c}{$4.153$}\\\hline
\multicolumn{6}{c}{$n=500$}\\\hline
BLM & $-0.098$ & \multicolumn{1}{r}{$0.411$} & $0.114$ & $0.324$ &
\multicolumn{1}{c}{$5.010$}\\
$\left(  \tau,\rho\right)  $-inversion & $-0.235$ & \multicolumn{1}{r}{$0.502$%
} & $0.136$ & $0.503$ & \multicolumn{1}{c}{$7.149$}\\
MD & $-0.181$ & \multicolumn{1}{r}{$0.409$} & $0.116$ & $0.376$ &
\multicolumn{1}{c}{$14.984$}\\
PML & $0.170$ & \multicolumn{1}{r}{$0.205$} & $-0.115$ & $0.619$ &
\multicolumn{1}{c}{$13.147$}\\\hline\hline
&  &  &  &  &
\end{tabular}
\ \ \ \ \ \ \ \ $%
\caption{Bias and RMSE of the BLM, ($\tau$,$\rho$)-inversion, MD and PML
estimators for two-parameters of Gumbel copula for weak dependence ($\rho=0.001$).}\label{A6}%
\end{table}%
}

{\tiny
\begin{table}[h] \centering
$%
\begin{tabular}
[c]{lrlrrl}
& \multicolumn{2}{c}{$\beta_{1}=1.4$} & \multicolumn{2}{c}{$\beta_{2}=0.2$} &
\\\hline
\ \ \ \ \ \ \ \ \ \ \ \ \ \ \ \  & \ \ \ \ \ \ \ Bias & \ \ \ \ \ RMSE &
\ \ \ \ \ \ Bias & \ \ RMSE & {\tiny Time (h)}\\\hline\hline
\multicolumn{6}{c}{$n=30$}\\\hline
BLM & $-0.182$ & \multicolumn{1}{r}{$0.989$} & $0.593$ & $1.317$ &
\multicolumn{1}{c}{$1.024$}\\
$\left(  \tau,\rho\right)  $-inversion & $0.191$ & \multicolumn{1}{r}{$0.885$}
& $0.655$ & $1.215$ & \multicolumn{1}{c}{$2.042$}\\
MD & $0.214$ & \multicolumn{1}{r}{$0.985$} & $-0.525$ & $1.056$ &
\multicolumn{1}{c}{$4.485$}\\
PML & $0.195$ & \multicolumn{1}{r}{$0.524$} & $-0.423$ & $1.051$ &
\multicolumn{1}{c}{$2.125$}\\\hline
\multicolumn{6}{c}{$n=50$}\\\hline
BLM & $-0.134$ & \multicolumn{1}{r}{$0.594$} & $0.526$ & $0.994$ &
\multicolumn{1}{c}{$1.058$}\\
$\left(  \tau,\rho\right)  $-inversion & $0.187$ & \multicolumn{1}{r}{$0.512$}
& $0.555$ & $0.972$ & \multicolumn{1}{c}{$2.212$}\\
MD & $0.181$ & \multicolumn{1}{r}{$0.423$} & $-0.461$ & $0.853$ &
\multicolumn{1}{c}{$5.588$}\\
PML & $0.177$ & \multicolumn{1}{r}{$0.318$} & $-0.413$ & $0.916$ &
\multicolumn{1}{c}{$2.859$}\\\hline
\multicolumn{6}{c}{$n=100$}\\\hline
BLM & $-0.122$ & \multicolumn{1}{r}{$0.482$} & $0.272$ & $0.492$ &
\multicolumn{1}{c}{$2.247$}\\
$\left(  \tau,\rho\right)  $-inversion & $-0.150$ & \multicolumn{1}{r}{$0.421$%
} & $0.291$ & $0.712$ & \multicolumn{1}{c}{$3.153$}\\
MD & $-0.170$ & \multicolumn{1}{r}{$0.439$} & $-0.269$ & $0.474$ &
\multicolumn{1}{c}{$6.256$}\\
PML & $0.152$ & \multicolumn{1}{r}{$0.293$} & $-0.275$ & $0.471$ &
\multicolumn{1}{c}{$5.254$}\\\hline
\multicolumn{6}{c}{$n=500$}\\\hline
BLM & $-0.101$ & \multicolumn{1}{r}{$0.223$} & $0.135$ & $0.312$ &
\multicolumn{1}{c}{$6.587$}\\
$\left(  \tau,\rho\right)  $-inversion & $-0.149$ & \multicolumn{1}{r}{$0.400$%
} & $0.221$ & $0.655$ & \multicolumn{1}{c}{$9.145$}\\
MD & $-0.106$ & \multicolumn{1}{r}{$0.306$} & $-0.212$ & $0.355$ &
\multicolumn{1}{c}{$14.445$}\\
PML & $0.102$ & \multicolumn{1}{r}{$0.221$} & $-0.200$ & $0.317$ &
\multicolumn{1}{c}{$13.157$}\\\hline\hline
&  &  &  &  &
\end{tabular}
\ \ \ \ \ \ \ \ $%
\caption{Bias and RMSE of the BLM, ($\tau$,$\rho$)-inversion, MD and PML
estimators for two-parameters of Gumbel copula for moderate dependence ($\rho=0.5$).}\label{A7}%
\end{table}%
}

{\tiny
\begin{table}[h] \centering
$%
\begin{tabular}
[c]{lrlrrc}
& \multicolumn{2}{c}{$\beta_{1}=2.5$} & \multicolumn{2}{c}{$\beta_{2}=1$} &
\\\hline
\ \ \ \ \ \ \ \ \ \ \ \ \ \ \ \  & \ \ \ \ \ \ \ Bias & \ \ \ \ \ RMSE &
\ \ \ \ \ \ Bias & \ \ RMSE & {\tiny Time (h)}\\\hline
\multicolumn{6}{c}{$n=30$}\\\hline\hline
BLM & $0.422$ & \multicolumn{1}{r}{$0.954$} & $-0.740$ & $1.119$ & $0.755$\\
$\left(  \tau,\rho\right)  $-inversion & $0.786$ & \multicolumn{1}{r}{$1.125$}
& $0.782$ & $1.175$ & $1.254$\\
MD & $0.546$ & \multicolumn{1}{r}{$0.546$} & $0.592$ & $0.563$ & $1.545$\\
PML & $0.553$ & \multicolumn{1}{r}{$0.551$} & $0.723$ & $0.522$ &
$1.765$\\\hline
\multicolumn{6}{c}{$n=50$}\\\hline
BLM & $0.329$ & \multicolumn{1}{r}{$0.817$} & $-0.635$ & $0.852$ & $1.021$\\
$\left(  \tau,\rho\right)  $-inversion & $0.586$ & \multicolumn{1}{r}{$0.983$}
& $0.745$ & $0.972$ & $2.045$\\
MD & $0.321$ & \multicolumn{1}{r}{$0.522$} & $0.582$ & $0.552$ & $2.265$\\
PML & $0.292$ & \multicolumn{1}{r}{$0.512$} & $0.551$ & $0.514$ &
$2.255$\\\hline
\multicolumn{6}{c}{$n=100$}\\\hline
BLM & $0.107$ & \multicolumn{1}{r}{$0.584$} & $-0.592$ & $0.713$ & $1.920$\\
$\left(  \tau,\rho\right)  $-inversion & $0.425$ & \multicolumn{1}{r}{$0.812$}
& $0.611$ & $0.902$ & $2.153$\\
MD & $-0.181$ & \multicolumn{1}{r}{$0.501$} & $-0.578$ & $0.488$ & $5.544$\\
PML & $0.172$ & \multicolumn{1}{r}{$0.482$} & $-0.545$ & $0.472$ &
$5.458$\\\hline
\multicolumn{6}{c}{$n=500$}\\\hline
BLM & $-0.066$ & \multicolumn{1}{r}{$0.456$} & $-0.367$ & $0.478$ & $5.205$\\
$\left(  \tau,\rho\right)  $-inversion & $0.123$ & \multicolumn{1}{r}{$0.757$}
& $0.501$ & $0.694$ & $9.789$\\
MD & $0.094$ & \multicolumn{1}{r}{$0.469$} & $0.408$ & $0.495$ & $14.565$\\
PML & $0.084$ & \multicolumn{1}{r}{$0.465$} & $0.375$ & $0.482$ &
$13.425$\\\hline\hline
&  &  &  &  &
\end{tabular}
\ \ \ \ \ \ \ \ \ \ $%
\caption{Bias and RMSE of the BLM, ($\tau$,$\rho$)-inversion, MD and PML
estimators for two parameters of Gumbel copula for strong dependence ($\rho=0.9$).}\label{A8}%
\end{table}%
}

{\tiny
\begin{table}[h] \centering
$%
\begin{tabular}
[c]{ccccc}
& \multicolumn{2}{c}{$\alpha_{1}=0.4$} & \multicolumn{2}{c}{$\alpha_{2}=0.9$%
}\\\hline
\multicolumn{5}{c}{$0\%$\ contamination}\\\hline
\ \ \ \ \ \ \ \ \ \ \ \ \ \ \ \  & \ \ \ \ \ \ \ Bias & \ \ \ \ \ RMSE &
\ \ \ \ \ \ Bias & \ \ RMSE\\\hline\hline
\multicolumn{1}{l}{BLM} & \multicolumn{1}{r}{$0.044$} &
\multicolumn{1}{r}{$0.832$} & \multicolumn{1}{r}{$-0.141$} &
\multicolumn{1}{r}{$2.650$}\\
\multicolumn{1}{l}{$\left(  \tau,\rho\right)  $-inversion} &
\multicolumn{1}{r}{$0.053$} & \multicolumn{1}{r}{$0.432$} &
\multicolumn{1}{r}{$-0.440$} & \multicolumn{1}{r}{$0.711$}\\
\multicolumn{1}{l}{MD} & \multicolumn{1}{r}{$0.267$} &
\multicolumn{1}{r}{$0.270$} & \multicolumn{1}{r}{$-0.456$} &
\multicolumn{1}{r}{$0.461$}\\
\multicolumn{1}{l}{PML} & \multicolumn{1}{r}{$-0.238$} &
\multicolumn{1}{r}{$0.440$} & \multicolumn{1}{r}{$-0.472$} &
\multicolumn{1}{r}{$0.627$}\\\hline
\multicolumn{5}{c}{$5\%$\ contamination}\\\hline
\multicolumn{1}{l}{BLM} & \multicolumn{1}{r}{$0.046$} &
\multicolumn{1}{r}{$0.833$} & \multicolumn{1}{r}{$-0.137$} &
\multicolumn{1}{r}{$2.662$}\\
\multicolumn{1}{l}{$\left(  \tau,\rho\right)  $-inversion} &
\multicolumn{1}{r}{$-0.069$} & \multicolumn{1}{r}{$0.431$} &
\multicolumn{1}{r}{$-0.479$} & \multicolumn{1}{r}{$0.738$}\\
\multicolumn{1}{l}{MD} & \multicolumn{1}{r}{$0.254$} &
\multicolumn{1}{r}{$0.257$} & \multicolumn{1}{r}{$-0.472$} &
\multicolumn{1}{r}{$0.475$}\\
\multicolumn{1}{l}{PML} & \multicolumn{1}{r}{$-0.274$} &
\multicolumn{1}{r}{$0.407$} & \multicolumn{1}{r}{$0.432$} &
\multicolumn{1}{r}{$0.613$}\\\hline
\multicolumn{5}{c}{$10\%$\ contamination}\\\hline
\multicolumn{1}{l}{BLM} & \multicolumn{1}{r}{$-0.082$} &
\multicolumn{1}{r}{$0.811$} & \multicolumn{1}{r}{$-0.155$} &
\multicolumn{1}{r}{$2.641$}\\
\multicolumn{1}{l}{$\left(  \tau,\rho\right)  $-inversion} &
\multicolumn{1}{r}{$-0.090$} & \multicolumn{1}{r}{$0.393$} &
\multicolumn{1}{r}{$-0.461$} & \multicolumn{1}{r}{$0.695$}\\
\multicolumn{1}{l}{MD} & \multicolumn{1}{r}{$0.279$} &
\multicolumn{1}{r}{$0.281$} & \multicolumn{1}{r}{$-0.464$} &
\multicolumn{1}{r}{$0.468$}\\
\multicolumn{1}{l}{PML} & \multicolumn{1}{r}{$-0.267$} &
\multicolumn{1}{r}{$0.506$} & \multicolumn{1}{r}{$-0.429$} &
\multicolumn{1}{r}{$0.637$}\\\hline
\multicolumn{5}{c}{$20\%$ contamination}\\\hline
\multicolumn{1}{l}{BLM} & \multicolumn{1}{r}{$-0.100$} &
\multicolumn{1}{r}{$0.802$} & \multicolumn{1}{r}{$-0.188$} &
\multicolumn{1}{r}{$2.585$}\\
\multicolumn{1}{l}{$\left(  \tau,\rho\right)  $-inversion} &
\multicolumn{1}{r}{$-0.130$} & \multicolumn{1}{r}{$0.423$} &
\multicolumn{1}{r}{$-0.537$} & \multicolumn{1}{r}{$0.786$}\\
\multicolumn{1}{l}{MD} & \multicolumn{1}{r}{$0.280$} &
\multicolumn{1}{r}{$0.282$} & \multicolumn{1}{r}{$-0.472$} &
\multicolumn{1}{r}{$0.477$}\\
\multicolumn{1}{l}{PML} & \multicolumn{1}{r}{$-0.268$} &
\multicolumn{1}{r}{$0.524$} & \multicolumn{1}{r}{$-0.500$} &
\multicolumn{1}{r}{$0.639$}\\\hline
\multicolumn{5}{c}{$30\%$\ contamination}\\\hline
\multicolumn{1}{l}{BLM} & \multicolumn{1}{r}{$-0.165$} &
\multicolumn{1}{r}{$0.835$} & \multicolumn{1}{r}{$-0.280$} &
\multicolumn{1}{r}{$2.627$}\\
\multicolumn{1}{l}{$\left(  \tau,\rho\right)  $-inversion} &
\multicolumn{1}{r}{$-0.179$} & \multicolumn{1}{r}{$0.480$} &
\multicolumn{1}{r}{$-0.619$} & \multicolumn{1}{r}{$0.909$}\\
\multicolumn{1}{l}{MD} & \multicolumn{1}{r}{$0.293$} &
\multicolumn{1}{r}{$0.266$} & \multicolumn{1}{r}{$-0.458$} &
\multicolumn{1}{r}{$0.465$}\\
\multicolumn{1}{l}{PML} & \multicolumn{1}{r}{$-0.328$} &
\multicolumn{1}{r}{$0.589$} & \multicolumn{1}{r}{$-0.515$} &
\multicolumn{1}{r}{$0.641$}\\\hline\hline
&  &  &  &
\end{tabular}
\ \ \ \ \ \ \ \ $%
\caption{Bias and RMSE of the BLM, ($\tau$,$\rho$)-inversion, MD and PML
estimators for $\epsilon$-contaminated two-parameters of FGM copula by product copula.}\label{A9}%
\end{table}%
}

\end{document}